\documentclass{jfm}

\usepackage{dsfont}
\usepackage{graphicx}
\usepackage{newtxtext}
\usepackage{newtxmath}
\usepackage{natbib}
\usepackage{hyperref}
\hypersetup{
	colorlinks = true,
	urlcolor   = blue,
	citecolor  = black,
}

\usepackage{csquotes}

\usepackage{booktabs}

\usepackage{pgfplots}
\pgfplotsset{compat=newest}
\usepackage{tikz}
\definecolor{mycolour}{rgb}{0,0.4470,0.7410}
\definecolor{colour1}{rgb}{0,0.4470,0.7410}
\definecolor{colour2}{rgb}{0.8500,0.3250,0.0980}
\definecolor{colour3}{rgb}{0.9290,0.6940,0.1250}
\definecolor{colour4}{rgb}{0.4940,0.1840,0.5560}
\definecolor{colour5}{rgb}{0.4660,0.6740,0.1880}
\definecolor{colour6}{rgb}{0.9,0.5,0}

\usetikzlibrary{shapes,arrows}
\usetikzlibrary{arrows.meta}

\tikzset{
	double arrow/.style args={#1 colored by #2 and #3}{
		-{Triangle[scale=0.5]},line width=#1,#2, 
		postaction={draw,-{Triangle[scale=0.5]},#3,line width=(#1)/3,
			shorten <=(#1)/3,shorten >=2*(#1)/3}, 
	}
}
\tikzset{
	fold arrow/.style args={}{
		double arrow=1.5pt colored by black and white
	}
}

\tikzset{
	sliding arrow/.style args={}{
		double arrow=1.5pt colored by black and red
	}
}

\usepackage{booktabs}
\usepackage{multirow}

\newcommand{\RomanNumeralCaps}[1]
\linenumbers

\title{Mathematical models and time-frequency heat maps for surface gravity waves generated by thin ships}

\author[N. R. Buttle et. al]{Nicholas R. Buttle\aff{1}, Ravindra Pethiyagoda\aff{1}, Timothy J. Moroney\aff{1}, Brian Winship\aff{2}, Gregor J. Macfarlane\aff{2}, Jonathan R. Binns\aff{3}
	\and~Scott W. McCue\aff{1}\corresp{\email{scott.mccue@qut.edu.au}}}

\affiliation{\aff{1}School of Mathematical Sciences, Queensland University of Technology, Brisbane QLD 4001, Australia
	\aff{2}Australian Maritime College, University of Tasmania, Launceston TAS 7250, Australia
	\aff{3}Defence Science and Technology Group, Fishermans Bend VIC 3207, Australia
}

\begin{document}
	
	\maketitle
	
\begin{abstract}
		
		Recent research suggests that studying the time-frequency response of ship wave signals has potential to shed light on a range of applications, such as inferring the dynamical and geometric properties of a moving vessel based on the surface elevation data detected at a single point in space. We continue this line of research here with a study of mathematical models for thin ships using standard Wigley hulls and Wigley transom-stern hulls as examples. Mathematical models of varying sophistication are considered, including basic minimal models, Michell's thin ship theory and the Hogner model, the latter two of which explicitly take into account the shape of the hull.
		We outline a methodology for carefully choosing the form and parameter values in the minimal models such that they reproduce the key features of the more complicated models in the time-frequency domain. For example, we find that a two-pressure model is capable of producing wave elevation signals that have a similar time-frequency profile as that for Michell's thin ship theory applied to the Wigley hull, including the crucially important features caused by interference between waves created at the bow and stern of the ship. One of the key tools in our analysis is the spectrogram, which is a heat-map visualisation in the time-frequency domain.  Our work here extends the existing knowledge on the topic of spectrograms of ship waves. The theoretical results in this paper are supported by experimental data collected in a towing tank at the Australian Maritime College using model versions of the standard Wigley hulls and Wigley transom-stern hulls.
\end{abstract}
	
	
\section{Introduction}
	\label{sec:intro}
	
We are concerned here with fundamental mathematical models for disturbances that generate ship waves, with particular reference to thin hull shapes and their corresponding time-frequency signatures.  The modelling framework here relies on linear water wave theory, with its rich history and well-known advantages in terms of being analytically tractable, and disadvantages for ignoring viscous dissipation and turbulent boundary layers \citep{Newman2018,Wehausen1973}.  Within that framework, one of our objectives here is to investigate necessary requirements for basic minimal models, especially in terms of generating realistic ship wakes that have relevant time-frequency signatures encoded within them, such as those that arise from wave interference.  As well as these minimal models, our study will involve Michell's thin ship approximation \citep{Michell1898,Tuck1989,Tuck2001} and the Hogner model \citep{Noblesse83,Zhu2017a,Zhu2018,liang2024} as practical but relatively simple theories; other related approaches are used in the literature, such as the more sophisticated Neumann-Kelvin theory \citep{Brard1972,Doctors1987}, or even the recently proposed Neumann-Michell model \citep{Huang2013,Noblesse2013,Wu2018a}.
	
\begin{figure}
		\centering
		\includegraphics[scale=1]{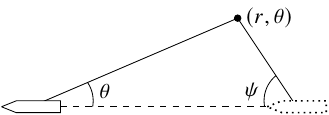}
		\caption{Schematic illustrating the meaning of $\theta$ and $\psi$.  The radial distance $r$ is measured from the centre of the ship hull, while $\theta$ is measured from the centreplane as indicated.  For a given point $(r,\theta)$ in the ship's wake, $\psi$ measures the angle at which waves have travelled to the point $(r,\theta)$ from where the ship was in the past.}
		\label{fig:schematic}
	\end{figure}

	Note that for all of the models mentioned, under the infinite-depth assumption, the far-field wave pattern is of the form
	\begin{align}
		\zeta(x,y) &\sim \int_{-\pi/2}^{\pi/2} A(\psi) \, \mathrm{e}^{-\mathrm{i} k_0 (x \cos \psi + y \sin \psi)} \, \mathrm{d} \psi \nonumber \\
		&= \int_{-\pi/2}^{\pi/2} A(\psi) \, \mathrm{e}^{-\mathrm{i} r g(\psi; \theta)} \, \mathrm{d} \psi, \label{eq:stationary_phase_version}
	\end{align}
	where the point $(x,y)$ in Cartesian coordinates, or $(r,\theta)$ in polar coordinates, is fixed in the reference frame of the moving ship.  In this representation, the details of the wave amplitude function $A(\psi)$ depend on the model chosen and
	\begin{align}
		g(\psi; \theta) &= k_{0} \cos(\psi + \theta). \label{eq:g_eq}
	\end{align}
	For the dimensionless version, $k_{0} = \sec^{2}\psi/ F^{2}$, where
	\begin{equation}
		F=\frac{U}{\sqrt{gL}}
		\label{eq:froude}
	\end{equation}
	is the length-based Froude number, $U$ is the speed of the ship and $L$ is the length of the ship hull.  For a geometric interpretation of $\theta$ and $\psi$, see figure~\ref{fig:schematic}. Waves that arrive at the point $(r,\theta)$ have propagated out from where the ship was in the past at an angle $\psi$ as indicated.
	
\begin{figure}
		\centering
		\begin{tabular}{ll}
			(a) Standard Wigley & (b) Wigley transom-stern \\
			\includegraphics[scale=1]{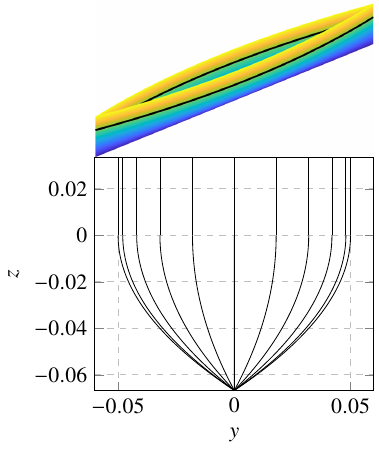} &
			\includegraphics[scale=1]{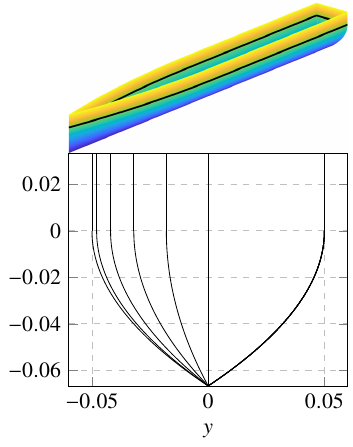} \\
		\end{tabular}
		\caption[Bodyline diagram]{Bodyline diagram for the two hulls used in our investigations with stations taken at $0.1$ units spacing. The two hulls are: (a) standard Wigley hull defined in \eqref{eq:Wigley} with nondimensional length $=1$, beam $\beta=0.1$, and draft $\delta=0.0667$; and (b) standard Wigley hull with transom-stern as defined in \eqref{eq:Wigley_transom_stern} with the same dimensions. The waterline is marked by the black line.}
		\label{fig:bodyline_diagrams}
	\end{figure}

	The two test cases we use involve the Wigley hull, which is parabolic in shape both longitudinally and laterally, and the Wigley transom-stern hull, which is the same as the standard Wigley from bow to midship and then subsequently has a constant profile from midship to stern. Figure~\ref{fig:bodyline_diagrams} shows bodyline diagrams and three-dimensional plots of both hulls with the waterline marked in black.  To study these test cases, we employ mathematical models with increasing complexity.  These include three very basic minimal models that involve Gaussian pressure distributions applied to the water's surface and point sources submerged below the surface. The use of applied pressure distributions to generate ship-like waves is widespread. 
As mentioned above, we also employ more sophisticated models that take the hull geometry into account, namely the Michell thin-ship model and the Hogner model.
	
	One of the goals of the study is to explore to what extent the minimal models with pressure distributions reproduce key features of the ship wake.  A part of this approach involves devising a methodology for how to best choose the parameters in the minimal models.  For example, two-point and four-point pressure models have been used in the past to study ship wakes \citep{colen21,He2015,Li2018,Noblesse2014,Pethiyagoda2018,Zhu2015,Zhu2018d}, and in this spirit one question is how far apart these localised pressures should be located.
	
	We pay particular attention to the interference between waves created at the bow and stern of a ship hull.  This is a topic of significant interest, especially in the context of linear water wave theory \citep{He2016,Noblesse2014,Zhang2015a,Zhu2018d}.  When calibrating parameters in our minimal pressure models, we shall endeavour to match certain properties of the relevant far-field wave amplitude function $A(\psi)$ in (\ref{eq:stationary_phase_version}) with those associated with the Michell or Hogner models.  In this way, we argue that wave interference in the wake of a ship hulls can be taken into account by focusing on the role of the wave amplitude function and not necessarily the wave pattern itself.
	
	A crucial tool we use in analysing our ship wakes is the spectrogram, which is a heat map in the time-frequency domain that visualises what wave frequencies are prominent at a certain point in space as the ship travels by \citep{Pethiyagoda2017,Torsvik2015}.  In particular, we utilise a wave elevation signal from a fixed sensor and apply short-time Fourier transforms to the signal to extract the frequencies of interest.  This type of spectrogram analysis is interesting as it has the potential for a number of significant applications, including: providing meaningful information about the root cause of the components of the wave pattern; to help track the cause of shoreline erosion in a shipping channel by attributing environmental damage to certain wave frequencies; and to facilitate the monitoring of unauthorised fishing vessels.  Further, importantly, the methodologies that arise from spectrogram analysis of ship wakes could in the future lead to tools to back-calculate information about ship hulls that create waves, such as the ship's location, its speed and hull shape. While in recent years some progress has been made on using spectrograms to study ship wakes and these applications \citep{Didenkulova2013,Liang2022,li2021,Luo2022,Pethiyagoda2017,Pethiyagoda2018,Pethiyagoda2021,Ratsep2020,Ratsep2020b,Ratsep2021,Safty2020,sheremet13,Torsvik2015}, no previous research has ever properly addressed how wave interference manifests itself in the time-frequency domain, nor how well minimal pressure models can be calibrated to produce very similar time-frequency responses to more complicated models that take the hull shape into account.
	
	The outline of the paper is as follows.  In \S\ref{sec:modellingframework} we outline the details for the basic minimal models with applied pressure distributions, Michell's thin ship model and the Hogner model, focusing on the amplitude function $A(\psi)$, including as much information that can be provided analytically.  Section~\ref{sec:wigley} focuses on the case of the Wigley hull.  We choose parameters for the two-pressure minimal model by matching properties of the amplitude function with that for Michell's thin ship theory.
	As such, we demonstrate how well the basic two-pressure model can mimic the more complicated thin ship theory in the time-frequency domain, especially in terms of the important wave interference that occurs for lower wave frequencies.
{In \S\ref{sec:transom} we apply a similar process for the Wigley transom-stern hull, except this time we extend the definition of the hull to add a virtual appendage behind stern, in an attempt to mimic flow separation that occurs due to the steep geometry of the transom stern.}
To support our theoretical study, we have conducted experiments in a model test basin using a particular Wigley hull and a Wigley transom-stern hull, as we describe in \S\ref{sec:experimental}.  By measuring wave elevation at fixed sensors as the hulls travel past, we are able to generate spectrograms to compare with our simulations.  We investigate how wave interference can lead to significantly different time-frequency responses when the speed of the Wigley hull changes only by relatively small values.  Finally, we finish in \S\ref{sec:discussion} with a discussion and summary.
	
	\section{Modelling framework}\label{sec:modellingframework}
	
	\subsection{Minimal models with applied pressure distributions}\label{sec:miminal}
	
	We begin with a rudimentary strategy using a single Gaussian pressure distribution to model the hull, moving at a constant velocity across an otherwise stationary fluid. We take the reference frame to move with the pressure distribution, which is equivalent to a stationary pressure distribution acting on a fluid with constant velocity in the far field. Standard linear water wave theory leads to Laplace's equation for the dimensionless velocity potential, $\phi(x,y,z)$, and the free-surface location, $z=\zeta(x,y)$:
	\begin{align}
		\nabla^{2} \phi &= 0, \quad z < 0, \label{eq:pressure_laplace}
	\end{align}
	with the dynamic and kinematic boundary conditions
	\begin{align}
		\phi_x + \frac{\zeta}{F^{2}} + \epsilon p &= 1 \quad \mathrm{on}\; z = 0, \label{eq:pressure_gov1}\\
		\zeta_{x} &= \phi_{z} \quad \mathrm{on}\; z = 0, \label{eq:pressure_gov2}
	\end{align}
	where, in this most simple case, $p(x,y) = \exp(-\pi^{2}(x^2 + y^2) / \sigma^{2})$ is a dimensionless axially symmetric Gaussian pressure distribution applied to the surface centred at the origin with ``half-width'' $\sigma$, $\epsilon$ is the dimensionless strength of the pressure and $F$ is the Froude number (\ref{eq:froude}). For this single pressure model, $L$ is left as an unspecified length-scale (which we interpret to be the length of a ship hull). We also impose the radiation condition
	\begin{align}
		\phi \sim x &\quad \mathrm{as} \quad x \rightarrow -\infty, \label{eq:pressure_gov3}
	\end{align}
	which ensures that the fluid is uniform ahead of the moving pressure (moving with unit speed in the far-field limit).
	
	Solving \eqref{eq:pressure_laplace} with \eqref{eq:pressure_gov1}-\eqref{eq:pressure_gov3} using Fourier transforms gives the solution
	\begin{align}
		\zeta &= -\epsilon p(x, y) + \frac{\epsilon}{2\pi^{2}} \int_{-\pi/2}^{\pi/2} \int_{0}^{\infty} \frac{k^{2} \tilde{p}(k, \psi) \cos(k [|x| \cos \psi + y \sin \psi ])}{k - k_{0}} \, \mathrm{d}k \, \mathrm{d}\psi \nonumber \\
		& \quad {} - \frac{\epsilon H(x)}{\pi} \int_{-\pi/2}^{\pi/2} k_{0}^{2} \tilde{p}(k_{0}, \psi) \sin(k_{0} [x \cos \psi + y \sin \psi ]) \, \mathrm{d}\psi, \label{eq:pressure_full_solution}
	\end{align}
	where $\tilde{p}(k,\psi) = \sigma^{2} \exp( -\sigma^{2}k^{2} / (4\pi^{2})) / \pi$ is the Fourier transform of the pressure distribution \citep{wehausen60}.
	
	We will be focused on analysing wave elevation signals that are sufficiently far from the ship, which means we can restrict ourselves to the far-field component, specifically the single integral in $\psi$ in \eqref{eq:pressure_full_solution}. After some algebra, this single integral can be rewritten in the form of \eqref{eq:stationary_phase_version} with the corresponding wave amplitude function
	\begin{align} \label{eq:single_pressure_amplitude}
		A_{1}(\psi) &= -\frac{\mathrm{i} \epsilon \sigma^{2} \sec^{4}\psi}{\pi^{2}F^{4}} \exp\left( -\frac{\sigma^{2} \sec^{4} \psi}{4 \pi^{2}F^{4} } \right).
	\end{align}
	Note that $A_1$ is purely imaginary. Here we use the subscript 1 as this is a single pressure model.
	
	
	We also propose a more complicated model that uses two applied Gaussian pressure distributions positioned at $(-\ell/2, 0)$ and $(\ell/2, 0)$, where we fix $\ell < 1$. These two distributions are the same shape, size and strength as each other, and so we call this the symmetric two-pressure model. This configuration of localised pressures on the surface is designed to be more representative of the forces exhibited by a hull at its bow and stern than the very simple single Gaussian model (broadly following \cite{colen21,Li2018,Noblesse2014,Pethiyagoda2018,Zhu2015}). Some other studies have used a source-sink pair to model the effect of the hull, which is analogous to the pair of positive pressures that we use \citep{Zhu2018d}. The wave amplitude function for the symmetric two-pressure model is
	\begin{align} \label{eq:two_pressure_amplitude}
		A_{2}(\psi) &= -\frac{\mathrm{i}\epsilon \sigma^{2} \sec^{4} \psi}{\pi^{2}F^{4}} \exp\left( -\frac{\sigma^{2} \sec^{4} \psi}{4\pi^{2} F^{4}} \right) \cos\left(\frac{\ell \sec \psi}{2F^{2}} \right),
	\end{align}
	which is purely imaginary. We have chosen the strength of the pressures in the symmetric two-pressure model to be half the strength of the pressure in the single pressure model; this has the effect of the magnitude of the wave amplitude function for the single pressure, $A_1$, acting as an envelope for the magnitude of the symmetric two-pressure amplitude function $A_2$. The obvious difference between $A_1$ and $A_2$ is the presence of the cosine term in (\ref{eq:two_pressure_amplitude}) that depends on the distance between the pressures, which introduces oscillations to the amplitude function and means that $A_2$ vanishes at certain values of $\psi$.  This feature of the model is due to wave interference between the waves created by the two pressures.
	
\begin{figure}
		\centering
		\includegraphics[scale=1]{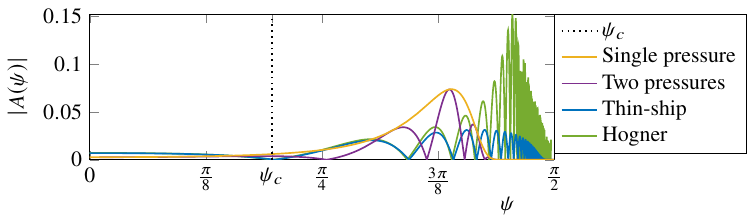}
		\caption{Magnitude of the wave amplitude functions for: the single-pressure model $A_1$ (\ref{eq:single_pressure_amplitude}) ({\color{colour3}orange}) and the symmetric two-pressure model $A_2$ (\ref{eq:two_pressure_amplitude}) ({\color{colour4}purple}) with $F = 0.3704$, $\epsilon = 0.05$, $\sigma = 0.1$, and $\ell = 0.9$; Michell's thin ship model $\left.A_{M}\right|_{W}$ (\ref{eq:thin_ship_amplitude})
			({\color{colour1}blue}) and the Hogner model $A_H$ (\ref{eq:hogner_amp}) ({\color{colour5}green}) for the standard Wigley hull \eqref{eq:Wigley} with $\beta = 0.1$, $\delta=0.067$ and $\nu = 0.0007$, again with $F = 0.3704$.}
		\label{fig:unmatched_amplitudes}
	\end{figure}

	As just mentioned, $A_{1}$ and $A_{2}$ are purely imaginary, as are all wave amplitude functions for bow-stern symmetric hulls. Figure \ref{fig:unmatched_amplitudes} shows representative amplitude functions (magnitude only) for the single-pressure ({\color{colour3}yellow}) and symmetric two-pressure ({\color{colour4}purple}) models (as well as Michell's thin-ship model and the Hogner model, which we will introduce later in this section). The parameter values for the pressure distributions were chosen to be: the half-width of the Gaussian pressure is the same as the beam of the hull, $\sigma=0.1$, the distance between the centres of the pressures is $90\%$ of the length of the hull, $\ell=0.9$, and the strength of the pressures is some small value, $\epsilon=0.05$.  All of these values are meant to be representative only; other more informed choices for $\sigma$, $\ell$ and $\epsilon$ will be discussed later in \S\ref{sec:pressure_estimation}. The particular Froude number used in this figure, $F=0.37$, was taken from an experimental run (with a hull of length $1.5\mathrm{m}$ and speed $1.42\mathrm{ms}^{-1}$), which will also be discussed later.  At this stage it is important to emphasise that the single pressure leads to a wave amplitude function whose magnitude is increasing up to a single maximum and then decreases again, while the symmetric two-pressure model has a number of local extrema and a sequence of zeros caused by wave interference.  As we shall show later, this difference has consequences in the time-frequency domain.
	
	{Motivated by the Wigley transom-stern hull, we also consider a more general minimal model consisting of two applied pressures, however now the longitudinal offsets, widths and strengths of the pressures are not necessarily connected to each other. The pressures are located at \((-\ell_{1}/2, 0)\) and \((\ell_{2}/2, 0)\) with widths \(\sigma_{1}\), \(\sigma_{2}\), and strengths \(\epsilon_{1}\), \(\epsilon_{2}\), respectively. The resulting more general wave amplitude function is:
	\begin{align}
		\left. A_{2} \right|_{T} (\psi) = -\frac{2\mathrm{i} \sec^{4} \psi}{\pi^{2}F^{4}}
		&\left[ \epsilon_{1} \sigma_{1}^{2} \exp\left( -\frac{\sigma_{1}^{2} \sec^{4} \psi}{4\pi^{2} F^{4}} \right) \exp\left(\frac{\ell_{1} \sec \psi}{2F^{2}} \right) \right. \nonumber \\
		&\left. {} + \epsilon_{2} \sigma_{2}^{2} \exp\left( -\frac{\sigma_{2}^{2} \sec^{4} \psi}{4\pi^{2} F^{4}} \right) \exp\left(-\frac{\ell_{2} \sec \psi}{2F^{2}} \right) \right], \label{eq:two_pressure_transom}
	\end{align}
where we use the notation $\left.A_{2} \right|_{T} (\psi)$ since there are two pressures and because we shall use this approach to model the transom-stern hull.	When \(\ell_{1} = \ell_{2}\), \(\sigma_{1} = \sigma_{2}\) and \(\epsilon_{1} = \epsilon_{2}\), then we recover the symmetric hull amplitude function (\ref{eq:two_pressure_amplitude}).}
	

	\subsection{Michell's thin ship model and the Hogner model}
	
	The next formulation considered is Michell's thin-ship theory \citep{Michell1898,Tuck1989,Tuck2001}. This approach attempts to simulate the effect of the ship hull using a distribution of sources projected onto the $y=0$ plane. When compared to our minimal models, this theory has the advantage of incorporating the particular geometry of the hull directly into the model.  This is done by setting the strength of the sources to be proportional to the $x$-derivative of the hull offset at that point, resulting in the double integral
	\begin{align}
		A_{M}(\psi) &= \frac{2\sec^{3} \psi}{\pi F^{2}} \int_{-\frac{1}{2}}^{\frac{1}{2}} \int_{-\delta}^{0} Y_{x}(x,z) \exp\left( \frac{z \sec^{2} \psi}{F^{2}} \right) \exp\left(\frac{\mathrm{i}\, x \sec \psi}{F^{2}} \right) \, \mathrm{d}z \, \mathrm{d}x,
		\label{eq:wigleymichell}
	\end{align}
	where $Y(x,z)$ is the shape of the hull in the positive $y$-direction, assuming the hull is laterally symmetric, and the subscript $M$ is used to denote the Michell theory. Using integration by parts, we can rewrite the solution in terms of the hull shape $y=Y(x,z)$. This allows us to compute wave amplitude functions of the form
	\begin{align}
		A_{M}(\psi) &= -\frac{2\mathrm{i}}{\pi F^4} \sec^{4}\psi \left( P(\psi) + \mathrm{i} Q(\psi) \right),
		\label{eq:amplitudeM}
	\end{align}
	where
	\begin{align}
		P(\psi) &= \int_{-\frac{1}{2}}^{\frac{1}{2}} \left[ \int_{-\delta}^{0} Y(x,z) \exp\left(\frac{z \sec^{2} \psi}{F^{2}}\right) \, \mathrm{d}z \right] \cos\left(\frac{x \sec \psi}{F^{2}}\right) \; \mathrm{d}x \nonumber \\
		& \quad {} - F^{2} \cos \psi \sin\left(\frac{\sec \psi}{2 F^2}\right)\int_{-\delta}^{0} Y(1/2,z) \exp\left(\frac{z \sec^{2} \psi}{F^{2}}\right) \, \mathrm{d}z,
		\label{eq:P}
		\\
		Q(\psi) &= \int_{-\frac{1}{2}}^{\frac{1}{2}} \left[ \int_{-\delta}^{0} Y(x,z) \exp\left(\frac{z \sec^{2} \psi}{F^{2}}\right) \, \mathrm{d}z \right] \sin\left(\frac{x \sec \psi}{F^{2}}\right) \; \mathrm{d}x \nonumber \\
		& \quad {} + F^{2} \cos \psi \cos\left(\frac{\sec \psi}{2 F^2}\right)\int_{-\delta}^{0} Y(1/2,z) \exp\left(\frac{z \sec^{2} \psi}{F^{2}}\right) \, \mathrm{d}z.
		\label{eq:Q}
	\end{align}
	For non-transom-stern hulls we have $Y(1/2,z)=0$, and therefore the second term in each of \eqref{eq:P} and \eqref{eq:Q} vanishes. For bow-stern symmetric hulls, $Q=0$, so that $A_{M}(\psi)$ is purely imaginary in that case.
	
	Unlike the Gaussian pressure models considered so far, the wave amplitude function $A_M(\psi)$ does not decay as $\psi \rightarrow \pi/2$, resulting in high-frequency waves that are not physically realistic and in reality would be damped out in water. This is a irksome feature of the Michell model.  To ensure that the amplitude function vanishes in the neighbourhood of $\psi =\pi/2$, we choose to limit the integral about the hull to extend up to a nondimensional distance $\nu$ below the surface.
	The integrals in \eqref{eq:P} and \eqref{eq:Q} in $z$ are now taken to be over the interval $z \in (-\delta, -\nu)$ instead of $z \in (-\delta, 0)$, where $\nu\ll 1$.
	
	There are at least three other options for damping out the high-frequency waves.  One of these, which can be applied to Michell's thin ship theory, is to include the effects of viscosity in the solution. This can be done by including a turbulent viscosity in the surface elevation integral \citep{Tuck2002} which damps out more of the high frequency waves as $r \rightarrow \infty$.  Another option is to alter Bernoulli's  equation on the surface to include viscous effects \citep{Cumberbatch1965,Liang2019}. A further option is to just \enquote{cut off} the amplitude functions at a certain value of $\psi$, which is essentially a crude version of the first method, although this approach results in numerical artefacts in the associated spectrograms.
	
	The last, and most complicated, theory for modelling thin ship hulls that we consider is the Hogner model, which is similar to Michell's thin-ship theory in that it represents the hull by sources that
	are proportional to the $x$-component of the hull normal; however, in the Hogner model the sources are now located at the surface of the actual hull, rather than along the $y=0$ plane. The far-field wave amplitude function for the Hogner model is defined by the double integral
	\begin{align}
		A_{H}(\psi) &= \frac{2 \sec^{3} \psi}{\pi F^{2}} \int_{-\frac{1}{2}}^{\frac{1}{2}} \int_{-\delta}^{0} \left[ Y_{x}(x,z) \exp\left( \frac{z \sec^{2} \psi}{F^{2}} \right) \exp\left( \frac{\mathrm{i}\, x \sec \psi}{F^{2}} \right) \right. \nonumber \\
		& \qquad \qquad \qquad \qquad \qquad \left. {} \times \cos\left( \frac{\sec \psi \tan \psi \,Y(x,z)}{F^{2}} \right) \right] \, \mathrm{d}z \, \mathrm{d}x. \label{eq:hogner_amp}
	\end{align}
	Unlike Michell's thin-ship model, the amplitude function for the Hogner model does decay as $\psi \rightarrow \pi/2$, so we integrate all the way up to the waterline without issue. We emphasise that, provided the hull can be represented $y=Y(x,z)$, the thin-ship model is a special limit of the Hogner model where the sources are projected onto the $y=0$ plane while retaining their strengths.
	
	\subsection{Wigley hulls and transom sterns}\label{sec:wigleytransom}
	
	For our investigations we will be considering a Wigley hull \citep{wigley1934} and a transom-stern variant, which are described mathematically as
	\begin{align}
		\text{Wigley hull:}& & Y(x,z) &= \pm \frac{\beta}{2} \left( 1 - \frac{z^{2}}{\delta^{2}} \right) \left(1 - 4x^{2} \right),
		\quad  -\frac{1}{2}<x<\frac{1}{2},
		\label{eq:Wigley}
		\\
		\text{Transom-stern:}& & Y(x,z) &= \pm
		\begin{cases}
			\displaystyle \frac{\beta}{2} \left( 1 - \frac{z^{2}}{\delta^{2}} \right) \left(1 - 4x^{2} \right), & -\dfrac{1}{2}<x < 0 \\
			\displaystyle \frac{\beta}{2} \left( 1 - \frac{z^{2}}{\delta^{2}} \right), & 0< x < \dfrac{1}{2}
		\end{cases},
		\\
		& & x &=\frac{1}{2}, \quad  |y|< \frac{\beta}{2} \left( 1 - \frac{z^{2}}{\delta^{2}} \right), &
		\label{eq:Wigley_transom_stern}
	\end{align}
	for $-\delta<z<0$, where the transom-stern variant is the same as the standard Wigley hull from the bow to the centre and is then a constant cross-section to the stern. Here, $\beta$ is the nondimensional beam of the hull and $\delta$ is the nondimensional draft, both having been scaled by the dimensional length of the hull. These two hull shapes are shown in figure \ref{fig:bodyline_diagrams}.
	
	It is important to note that these hulls are unstable in practice and so neither are seaworthy.  Instead, they are convenient from a modelling perspective as they are represented by the simple formulae in (\ref{eq:Wigley})-(\ref{eq:Wigley_transom_stern}), fully parameterised by the beam $\beta$ and draft $\delta$.  For this reason, Wigley hulls are used widely in the literature (see, for example \cite{chen83,kim07,liu2001,nakos90}).
	
	The wave amplitude function (\ref{eq:amplitudeM}) for Michell's thin-ship model can be derived analytically in the case of the Wigley hull as
	\begin{align}
		\left.A_{M}\right|_{W}(\psi) &= -\frac{8 F^{2} \beta \mathrm{i} \sqrt{1 + 4 F^{4} \cos^{2} \psi}}{\pi\delta^{2} \sec^{4} \psi}\cos \left( \arctan(2F^{2}\cos \psi) + \frac{\sec \psi}{2F^{2}} \right) \nonumber \\
		& \quad {} \times \left[ -\left( \nu + F^{2} \cos^{2} \psi \right)^{2} \exp\left( -\frac{\nu \sec^{2} \psi}{F^{2}} \right) \right. \nonumber \\
		& \qquad {} + \left( \delta + F^{2} \cos^{2} \psi \right)^{2} \exp\left( -\frac{\delta \sec^{2} \psi}{F^{2}} \right) \nonumber \\
		& \qquad {} \left. + \left( \delta^{2} - F^{4} \cos^{4} \psi \right) \left( \exp\left( -\frac{\nu \sec^{2} \psi}{F^{2}} \right) - \exp\left( -\frac{\delta \sec^{2} \psi}{F^{2}} \right) \right) \right]. \label{eq:thin_ship_amplitude}
	\end{align}
	The first two lines of \eqref{eq:thin_ship_amplitude} act as an envelope for the amplitude function, while the cosine is an oscillatory factor that introduces values of $\psi$ for which $\left.A_{M}\right|_{W}=0$.
	Note that the oscillatory factor depends only on the Froude number and not the beam and draft of the hull, indicating that the roots will be the same for all Wigley hulls with the same length travelling at the same speed. As mentioned above, since the standard Wigley hull is bow-stern symmetric, the wave amplitude function (\ref{eq:thin_ship_amplitude}) is purely imaginary.
	
	The wave amplitude for the Wigley hull using the Hogner model, $\left.A_{H}\right|_{W}$, does not simplify down analytically and must be computed numerically.  Given our Wigley hull is quite thin, it turns out that the Hogner model follows Michell's thin ship model closely, at least for small and moderate values of $\psi$ (they do not agree for $\psi$ closer to $\pi/2$, as discussed below).
	
	Both $\left.A_{M}\right|_{W}$ and $\left.A_{H}\right|_{W}$ are plotted in figure \ref{fig:unmatched_amplitudes}. The locations of zero amplitude for the thin-ship and Hogner models align very closely but are not exactly the same.  As mentioned above, these roots will match if we let the sources in the Hogner model keep their original strengths but project them to be on the $y=0$ plane, rather than be located on the geometry of the hull. In this way, the thin-ship model can be seen as a special limit of the Hogner model, for the case of an extremely thin ship. This also means that the roots of $A_H$ are dependent on the particular shape of the hull, unlike in Michell's thin-ship model.

For the transom-stern hull, a straightforward application of either Michell's thin ship theory or the Hogner model is deficient in the sense it does not take into account the significant flow separation and induced vorticity due to the very steep nature of the stern, which is clearly not at all ``thin'' (an assumption employed to derive the two thin ship models).  As an approximation, we follow previous researchers and suppose there is a virtual appendage behind the stern that encloses the separation zone.  In particular, the length of the separation zone is approximated to be $c$-times the width of the beam of the hull at each point of the transom stern and is defined as
	\begin{equation*}
		X_{v}(y, z) = x_{s} + 2cY(x_{s}, z) \frac{cY_{x}(x_{s}, z) + \sqrt{\left( c Y_{x}(x_{s}, z) + 1 \right)^{2} - \frac{2 c Y_{x}(x_{s}, z) + 1}{Y(x_{s}, z)}Y^{*}}}{2 c Y_{x}(x_{s}, z) + 1},
	\end{equation*}
	which we can simplify for our hull, using the fact that $Y_{x}(x_{s},z) = 0$, to
	\begin{equation*} \label{eq:virtual_appendage}
		X_{v}(y, z) = x_{s} + 2 c Y(x_{s}, z) \sqrt{1 - \frac{Y^{*}}{Y(x_{s}, z)}}.
	\end{equation*}
	We use $c = 3$ based on the study by \citet{Couser1998}. This method extends the hull past the transom-stern to account for the flow separation behind the hull.  Taking this vertical appendage into account, we denote the wave amplitude function due to Michell's thin ship theory and the Hogner model by $\left.A_{M}\right|_{T}(\psi)$ and $\left.A_{H}\right|_{T}(\psi)$, respectively.  Since in this case neither of these amplitude functions can be calculated exactly, we prefer the Hogner model for tuning with our simple two-pressure model (in section~\ref{sec:transom}) as it is more physically accurate.

	\subsection{Stationary-phase approximation}\label{sec:statphase}
	
\begin{figure}
		\centering
		\begin{tabular}{c}
			(a) $\theta$ vs $\psi$ \\
			\includegraphics[scale=0.8]{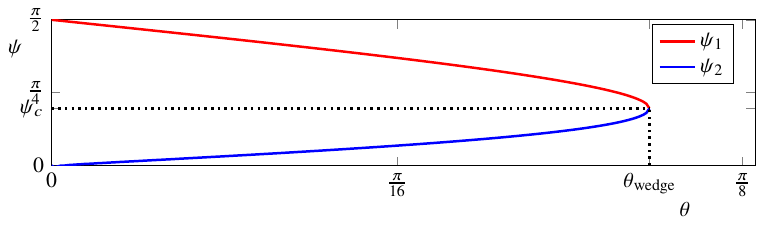} \\
			(b) weight function \\
			\includegraphics[scale=0.8]{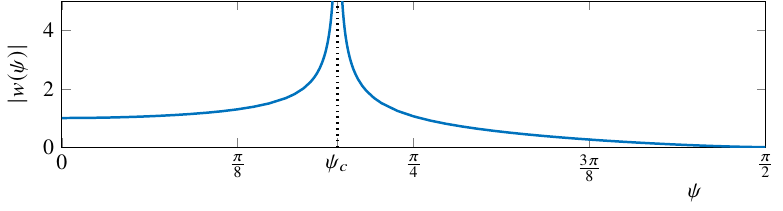}
		\end{tabular}
		\caption{(a) Plot of $\psi_1$ and $\psi_2$ from (\ref{eq:defnpsi}). These are the angles that provide the dominant stationary phase contributions. (b) Plot of the weight function \eqref{eq:uniform_expansion_weight} of the amplitude function in the stationary phase approximation \eqref{eq:stationary_phase} for the surface elevation.}
		\label{fig:weight}
	\end{figure}

	
	As is well known, we can approximate the surface elevation for our different models by utilising a stationary-phase approximation to \eqref{eq:stationary_phase_version}.  While the wave elevation is provided as an integral over all of $\psi$, the dominant contribution comes from the points of stationary phase, namely when $g'(\psi;\theta)=0$, where here the dash is used to mean a derivative with respect to $\psi$.  Provided $\theta<\theta_{\mathrm{wedge}}$, where $\theta_{\mathrm{wedge}}=\arcsin(1/3)=\arctan(1/\sqrt{8})\approx 19.47^\circ$ is the Kelvin wake angle, there are two points of stationary phase given by
	\begin{equation}
		\psi_1= \arctan\left(\frac{1+\sqrt{1-8\tan^2\theta}}{4\tan\theta}\right), \quad
		\psi_2= \arctan\left(\frac{1-\sqrt{1-8\tan^2\theta}}{4\tan\theta}\right),
		\label{eq:defnpsi}
	\end{equation}
	where $\psi_1$ and $\psi_2$ correspond to the divergent and transverse waves, respectively, as shown by figure~\ref{fig:weight}(a). The stationary phase approximation for $\theta<\theta_{\mathrm{wedge}}$ is
	\begin{align}
		\zeta(r,\theta) &\sim \Re\left\lbrace\frac{\sqrt{2\pi}}{r^{1/2}}\left( \frac{A(\psi_1)}{\sqrt{-g''(\psi_1)}}\exp\left(\mathrm{i}\left(rg(\psi_1)-\frac{\pi}{4}\right)\right) +\frac{A(\psi_2)}{\sqrt{g''(\psi_2)}}\exp\left(\mathrm{i}\left(rg(\psi_2)+\frac{\pi}{4}\right)\right)\right)\right\rbrace \label{eq:stationary_phase}
	\end{align}
	as $r \rightarrow \infty$.  Clearly (\ref{eq:stationary_phase}) shows this approximation at a point $(r,\theta)$ is a linear combination of a divergent and transverse wave arising from $\psi=\psi_1$ and $\psi_2$, respectively.  These are the two angles at which a monochromatic wave propagating at group velocity can travel from where a ship was in the past to the point $(r,\theta)$, as indicated in figure~\ref{fig:schematic} \citep{Stoker1957}.
	
	For our purposes, it is important to highlight the role of the wave amplitude function $A(\psi)$ in this approximation.  We see from (\ref{eq:defnpsi}) that for each $\theta<\theta_{\mathrm{wedge}}$ we have $0\leq\psi_2<\psi_c$ and $\psi_c<\psi_1<\pi/2$, where the critical angle $\psi_c=\arctan(1/\sqrt{2})$.  Thus for any ray angle $\theta$, the wave elevation depends crucially on the values of $A(\psi_1)$ and $A(\psi_2)$.  For example, if either $A(\psi_1)$ or $A(\psi_2)$ are precisely zero on a particular ray angle $\theta$, then either divergent or transverse waves do not appear (in this approximation), which is a consequence of destructive wave interference (they would appear at higher orders under a standard steepest descent expansion).
	
	Further, another observation from (\ref{eq:stationary_phase}) is the presence of the weighting function
	\begin{align}
		w(\psi) &= \sqrt{\frac{1}{\left| g''(\psi) \right|}}, \label{eq:uniform_expansion_weight}
	\end{align}
	which multiplies $A(\psi)$.  This function is singular at the angle $\psi_{c}$ (corresponding to $\theta=\theta_{\mathrm{wedge}}$) and decays as $\psi \rightarrow \pi/2$ (corresponding to divergent waves along the centreline $\theta=0$), as indicated by figure~\ref{fig:weight}(b).  Thus we see that the method of stationary phase approximation suggests that values of $\psi_1$ and $\psi_2$ close to $\psi_c$ are crucial in terms of the overall wave elevation, while the details of the wave amplitude functions $A(\psi)$ near $\psi=\pi/2$ are not important.
	
	Of course the method of stationary phase approximation does not work on (or very close to) the boundary of the Kelvin wedge as the two stationary points $\psi_1$ and $\psi_2$ collide when $\theta=\theta_{\mathrm{wedge}}$.  A better approximation is the uniform expansion of \eqref{eq:stationary_phase_version}, which is
	\begin{align}
		\zeta &\sim \Re\left\lbrace\frac{\sqrt{2}\pi\Delta^{1/4}}{r^{1/3}}\left(\frac{A(\psi_1)}{\sqrt{- g''(\psi_1)}}+ \frac{A(\psi_2)}{\sqrt{g''(\psi_2)}} \right)\exp\left(\mathrm{i}r\tilde{g}\right)\mathrm{Ai}(-r^{2/3}\Delta)\right.\\
		&\qquad\left.+\mathrm{i}\frac{\sqrt{2}\pi }{r^{2/3}\Delta^{1/4}}\left(\frac{A(\psi_1)}{\sqrt{- g''(\psi_1)}}- \frac{A(\psi_2)}{ \sqrt{g''(\psi_2)}}\right)\exp\left(\mathrm{i}r\tilde{g}\right)\mathrm{Ai}'(-r^{2/3}\Delta)\right\rbrace
		\label{eq:uniformexpansion}
	\end{align}
	as $r \rightarrow \infty$, where
	\begin{equation}
		\tilde{g} = \frac{1}{2}\left(g(\psi_1)+g(\psi_2)\right),
		\quad
		\Delta= \left(\frac{3}{4}(g(\psi_1)-g(\psi_2))\right)^{2/3},
		\nonumber
	\end{equation}
	and $\mathrm{Ai}(z)$ is the Airy function of the first kind \citep{Berry2010}.  Note the branches are chosen such that $\Delta$ is real positive when $\psi_{1,2}$ are real, and real negative otherwise.  In particular, for $\theta=\theta_{\mathrm{wedge}}$, this uniform expansion reduces to
	$$
	\zeta \sim \Re\left\lbrace
	\frac{\Gamma(1/3)}{\sqrt{3}r^{1/3}}\left(
	\frac{6}{g'''(\psi_c)}
	\right)^{1/3}
	A(\psi_c)
	\exp(\mathrm{i}rg(\psi_c))
	\right\rbrace
	\quad\mbox{as}\quad r\rightarrow\infty.
	$$
	The key point here is that even after correcting for the deficiencies in the stationary phase approximation near $\psi=\psi_c$, we still see that $A(\psi_1)$ and $A(\psi_2)$ are crucially important, when $\psi_1$ and $\psi_2$ are near $\psi_c$.
	Discussions and further details on stationary phase approximation and uniform expansions are found in \cite{chester57,Ursell1960,Warren1961}.
	
	\section{Parameter estimation and spectrograms for standard Wigley hull}\label{sec:wigley}
	\subsection{Parameter selection for minimal two-pressure model}\label{sec:pressure_estimation}
	
	Previous studies have shown that the minimal single pressure model, or similar basic applied pressure models, are capable of producing Kelvin wave patterns that incorporate all the key features of ship waves
	\citep{colen21,Darmon2014,Ellingsen2014,GabrielDavid2017,Li2002,Li2016,lo21,Miao2015,Parau2002,Pethiyagoda2015,Pethiyagoda2021a,Rabaud2014,Scullen2011,Shi2022,Smeltzer2019}.
	These properties include the shape and pattern of the divergent and transverse waves, including the manner in which the divergent waves become more prominent as the Froude number increases, as well as the effect that Froude number has on the apparent wake angle (see \cite{Darmon2014,dias14,Noblesse2014,Pethiyagoda2014a,Rabaud2013} for further details of these studies).  However, simple single-pressure models are unable to capture the more complicated behaviour that arises from constructive or destructive interactions between bow- and stern-generated waves or other related wave interference.
	
	This deficiency is part of the motivation for using a two-pressure model.  The other main motivation is to retain the simplicity and numerical efficiency that comes with a pressure model.  A challenge, therefore, is to bridge the gap between pressure models and more realistic theories that take into account the shape of a hull by carefully choosing parameter values in the two-pressure model.  To achieve this aim, the broad idea is to start with a particular hull, implement Michell's thin ship theory to calculate the wave amplitude function $A_M(\psi)$, choose the parameter values in $A_2(\psi)$ to mimic the key features of $A_M(\psi)$, and reflect on the match in terms of the time-frequency output via a spectrogram.
	
\begin{figure}
		\begin{tabular}{l}
			(a) $F = 0.287$ \\
			\includegraphics[width=0.9\textwidth]{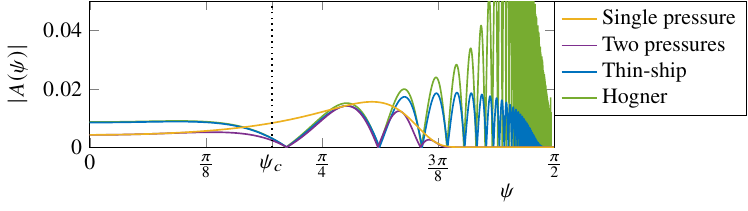} \\
			(b) $F = 0.334$ \\
			\includegraphics[width=0.9\textwidth]{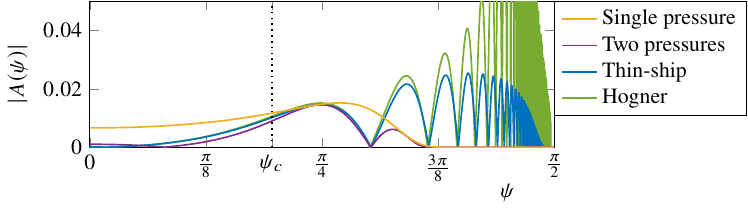} \\
			(c) $F = 0.370$ \\
			\includegraphics[width=0.9\textwidth]{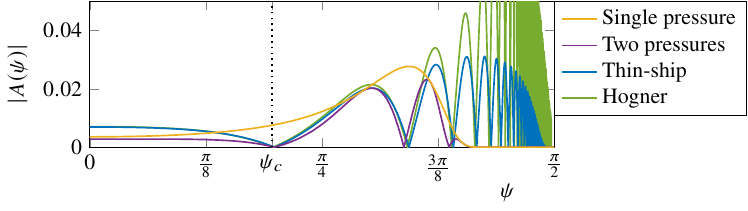}
		\end{tabular}
		\caption{Magnitude of the wave amplitude functions for the single-pressure model $A_1$ (\ref{eq:single_pressure_amplitude}) ({\color{colour3}orange}), the symmetric two-pressure model $A_2$ (\ref{eq:two_pressure_amplitude}) ({\color{colour4}purple}), Michell's thin ship model $\left.A_{M}\right|_{W}$ (\ref{eq:thin_ship_amplitude}) ({\color{colour1}blue}) and the Hogner model $A_H$ (\ref{eq:hogner_amp}) ({\color{colour5}green}), the latter two of which computed for the standard Wigley hull \eqref{eq:Wigley} with $\beta = 0.1$, $\delta=0.067$ and $\nu=0.01\delta$. (a) $F=0.287$, $\sigma=0.1729$, $\ell=1.0167$, $\epsilon = 0.0106$.  (b) $F=0.334$, $\sigma=0.3086$, $\ell=1.0167$, $\epsilon = 0.0103$.  (c) $F=0.370$, $\sigma=0.1915$, $\ell=1.0487$, $\epsilon = 0.0188$.}
		\label{fig:amplitudes}
	\end{figure}

	We now describe this approach in more detail.  Given a particular hull shape, the wave amplitude function for Michell's thin ship theory, $A_M(\psi)$, depends on the Froude number $F$.  Therefore we start by fixing the Froude number.  For example, for our Wigley hull illustrated in figure~\ref{fig:bodyline_diagrams}, the wave amplitude function (\ref{eq:thin_ship_amplitude}) can be computed (see the blue curves in figures~\ref{fig:unmatched_amplitudes} and \ref{fig:amplitudes}.  The next step is to choose the parameter $\ell$ in the symmetric two-pressure model by aligning a root of the wave amplitude function $A_2(\psi)$ with that for $A_M(\psi)$.  As we have argued above in \S\ref{sec:statphase}, the form of the amplitude function near $\psi=\psi_c$ is critically important for the overall wave pattern.  Therefore when it comes to the effects of wave interference, we expect an important feature of the amplitude function to mimic is the location of the root closest to $\psi=\psi_c$.  With this in mind, we determine the closest root of $A_M(\psi)$ to $\psi=\psi_c$, call it $\bar{\psi}$ say, and then choose $\ell$ so that $A_2(\psi)$ has a root at $\psi=\bar{\psi}$.  For the Wigley hull, using \eqref{eq:thin_ship_amplitude}, this process involves solving the nonlinear algebraic equation
	\begin{align*}
		0 &= \arctan\left( 2 F^{2} \cos \bar{\psi}\right) + \frac{\sec\bar{\psi}}{2 F^{2}} - \frac{(2n+1)\pi}{2}
	\end{align*}
	numerically, where $n\in \mathds{Z}$ is used to choose the root of $A_M$ closest to $\psi_c$.  With that value of $\bar{\psi}$, we then set
	\begin{align*}
		\ell &= (2 n + 1) \pi F^{2} \cos \overline{\psi},
	\end{align*}
	where $n$ ensures that $A_{M}$ and $A_{2}$ have the same number of roots in $0 \leq \psi \leq \bar{\psi}$.
	
	Once $\ell$ has been determined, the values for $\sigma$ and $\epsilon$ are chosen to match the location and height of the local maximum of the magnitude of the wave amplitude function that is closest to $\psi=\psi_c$.  The location is determined by differentiating \eqref{eq:thin_ship_amplitude} with respect to $\psi$ and then using a nonlinear solver to determine the value of $\psi$, $\psi^*$ say, at which the derivative is zero.  Once $\psi^*$ is determined, we set the value of $\sigma$ to be
	\begin{align*}
		\sigma &= \frac{\pi \sqrt{2} F}{2} \sqrt{\cos^{3} \psi^{*} \left( -\ell \tan \left( \frac{\ell \sec \psi^{*}}{2F^{2}} \right) + 8F^{2} \cos \psi^{*} \right) }.
	\end{align*}
	The value for $\epsilon$ is then chosen to match the magnitudes of the wave amplitude functions at $\psi^{*}$.
	
	The results of this parameter estimation are evident in figure~\ref{fig:amplitudes}, which shows wave amplitude functions for the four models.  Three cases are shown, namely (a) $F=0.287$, (b) $F=0.334$ and (c) $F=0.370$.  In all cases, the thin-ship model and the Hogner model are computed for the standard Wigley hull with $\beta=0.1$ and $\delta=0.067$.  The first important observation is that by following our algorithm, we are able to choose parameters for the symmetric two-pressure model so that the wave amplitude function for this model follows that for the thin-ship theory for a significant interval in $\psi$ around $\psi=\psi_c$. For example, in (a) these two curves are close to each other for roughly $\psi \in (\psi_{c}, 5\pi/16)$. As we shall see later, this close match will lead to the spectrograms for these models to share key features.
	Sticking with (a), we see that for roughly $\psi>5\pi/16$ there is no agreement between the two wave amplitude functions at all.  While these obvious differences will ultimately affect the corresponding spectrograms, we argue that these are unimportant as high frequency waves are damped out in real-life scenarios. As such, the summary at this point is that we can manipulate parameters in a minimal two-pressure model to mimic the more complicated thin ship theory for significant intervals in $\psi$.
	
	To test this idea further we include in figure~\ref{fig:amplitudes}$(b)$ a special case for which our methodology is most obviously deficient. Here, the Froude number $F=0.334$ is chosen so that $A_M(0)=0$.  That is, for this Wigley hull, Michell's thin ship theory predicts that the contribution to transverse waves along the centreline $\theta=0$ (which comes from $\psi=0$) is zero, due to exact destructive wave interference between transverse waves created at the bow and stern. To see this via the stationary phase approximation, see (\ref{eq:stationary_phase}).  In this special case $F=0.334$, our match between the symmetric two-pressure model and the thin-ship model is best for $\psi \in (\psi_{c}, 5\pi/16)$ and is close for $\psi \in (0, \psi_{c})$. While the first root in the amplitude function for the two-pressure model is close to $\psi = 0$, it is still some distance away leading to the transverse waves having a larger amplitude than in the thin-ship model. Since the transverse waves are very easy to identify on the spectrogram, this will cause obvious differences in their time-frequency plots. We shall reflect on the corresponding spectrograms in this case below.
	
	Finally, the third example in figure~\ref{fig:amplitudes} is for $F=0.370$, shown in $(c)$.  Here the Froude number is chosen so that the wave amplitude function for Michell's thin ship model has a zero at $\psi=\psi_c$.  Here destructive wave interference is such that the model predicts no wave contribution along the Kelvin wedge angle $\theta=\theta_{\mathrm{wedge}}$ (which comes from $\psi=\psi_c$; see the uniform expansion (\ref{eq:uniformexpansion})).  In this case our algorithm produces a very good match for much of the domain in a similar way to that shown in $(a)$.
	
	In each of the three panels in figure~\ref{fig:amplitudes} we also show the very simple single-pressure model and the Hogner model, the latter of which is the most complicated of our approaches.  The single-pressure wave amplitude function acts as an envelope for the symmetric two-pressure model but, apart from that feature, the match is not very good.  In this respect, we see that while they are both very basic models, the use of two pressures allows for much greater flexibility and promise than the single pressure, especially when it comes to matching with a more complicated model like Michell's thin theory.  Regarding the Hogner model, for this particular hull type, namely the standard Wigley hull, the Hogner model is certainly very close to the thin ship model for much of the domain, except for the less important values of $\psi$ near $\pi/2$. However, the Hogner model is analytically and computationally more challenging than the thin ship model, which makes the Hogner model less appealing for this hull shape, at least for our purposes. While we have used the thin-ship model to determine the parameters for the symmetric two-pressure model, we could have equally used the Hogner model. The thin-ship model has the advantage of having an exact form of the amplitude function for this hull shape, whereas extracting features from the Hogner model would involve finding the different features, namely the root and local maximum of interest, numerically.
	
\begin{figure}
		\centering
		\begin{tabular}{l}
			(a) Wigley hull \\
			\includegraphics[scale=1]{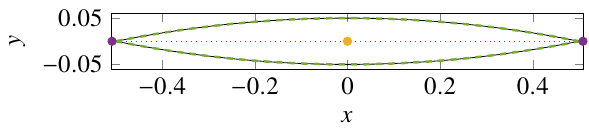} \\
			(b) Wigley transom-stern hull \\
			\includegraphics[scale=1]{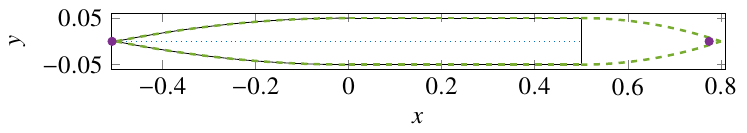}
		\end{tabular}
		\caption{Diagram of the locations of the pressure distributions for the Froude number $F = 0.287$.  (a) Wigley hull; pressure for the single-pressure model centred at the {\color{colour3}yellow} dot; pressures for the symmetric two-pressure model centred at the {\color{colour4}purple} dots; sources for the thin-ship model located along the centreline represented as a {\color{colour1}blue dotted} line; sources for the Hogner model located at hull itself, represented by the {\color{colour5}green dashed} curve.  (b) Wigley transom-stern hull; pressures for the general two-pressure model centred at the {\color{colour4}purple} dots; {\color{colour1}blue dotted} line and {\color{colour5}green dashed} curve represent the thin-ship model and the Hogner model, taking into account the virtual appendage behind the stern, designed to enclose the separation zone.}
		\label{fig:pressure_source_locations}
	\end{figure}

	Figure~\ref{fig:pressure_source_locations}(a) shows the locations of the pressure distributions and sources used by the different models discussed in figure~\ref{fig:amplitudes}.  We choose the $F=0.287$ from figure~\ref{fig:amplitudes}(a), but the results for the other two Froude numbers are similar.  The two minimal models have the pressure distributions centred at the nodes marked on the figure, while the thin-ship model has a distribution of sources along the $y=0$ plane and the Hogner model has a distribution of sources along the shape of the hull.  We shall return to figure~\ref{fig:pressure_source_locations}(b) below.
	
	\subsection{Time-frequency maps}
	
	In the previous section we outlined one method for choosing parameter values in a minimal two-pressure model so that its wave amplitude function follows that for Michell's thin ship theory applied to a standard Wigley hull, at least for certain important intervals in the wave angle $\psi$.  Our real motivation is to test how well a two-pressure model can mimic the more complicated Michell model in the time-frequency domain.  In particular, we study a wave elevation signal measured at a constant horizontal offset from the sailing line via a spectrogram, which shows the intensity of the different wave frequencies across time.  These signals are equivalent to those measured at a stationary sensor fixed in space as the ship travels by.
	
	To compute a spectrogram, we first need to generate a wave signal $s(t)$. In the laboratory reference frame, we measure the wave elevation at a sensor fixed in space as the ship travels past.  In the reference frame of the ship, the same signal is found by taking a cross-section of the wave profile at a constant value of $y$, $y_s$ say, and setting $s(t)=\zeta(t,y_s)$, given that the speed of the ship is unity in our dimensionless variables.  The spectrogram data is then given by the square magnitude of the short-time Fourier transform,
	\begin{align*}
		S(t,f) &= \left|\int_{-\infty}^{\infty}h(\tau-t)s(\tau)\mathrm{e}^{-2\pi\mathrm{i}f\tau}\, \mathrm{d}\tau\right|^2,
	\end{align*}
	\citep{sejdic09} where here $h(t)$ is a window function that is even with compact support \citep{Harris1978}.  The spectrogram itself is a heat-map visualisation of $S$, with frequency on the vertical axis, scaled time $x/y_s$ on the horizontal axis, and a colour intensity representing the logarithm of $S$ to the base 10.
	
	While various properties of spectrograms of ship waves have been studied, previous research \citep{Pethiyagoda2017,Torsvik2015a} demonstrates how linear water wave theory suggests there will be two main branches on the spectrogram, a constant frequency branch for the transverse waves and a sliding frequency branch for the divergent waves.  These essential properties are common to all of our spectrograms, regardless of the model used or the parameter values chosen (the effects of finite depth and ship acceleration lead to further complications, as explain in \cite{Pethiyagoda2018,Pethiyagoda2021}).
	
\begin{figure}
		\centering
		\begin{tabular}{ll}
			(a) Single-pressure model & (b) Symmetric two-pressure model \\
			\includegraphics[scale=1]{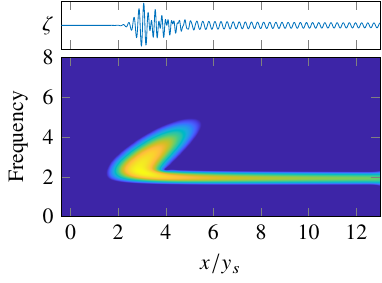} &
			\includegraphics[scale=1]{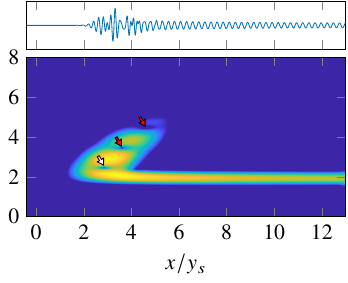} \\
			(c) Michell's thin-ship model & (d) Hogner model \\
			\includegraphics[scale=1]{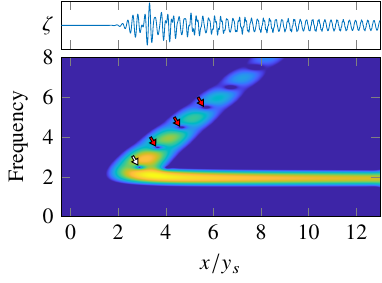} &
			\includegraphics[scale=1]{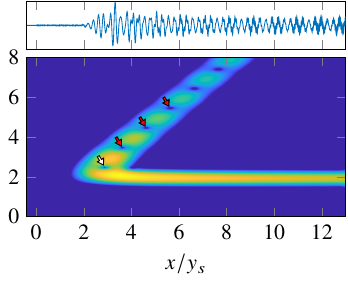} \\
			(e) Experimental standard hull size& (f) Experimental small hull \\
			\includegraphics[scale=1]{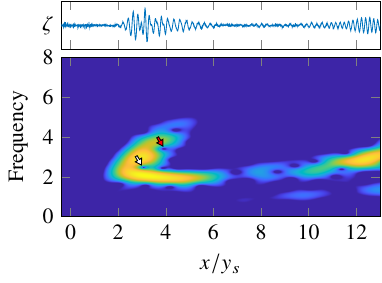} &
			\includegraphics[scale=1]{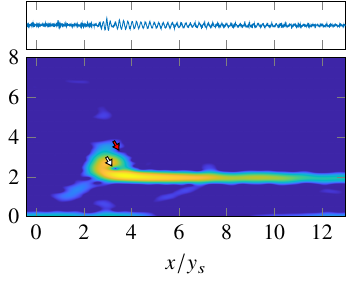}
		\end{tabular}
		\caption{Spectrograms for the (a) single-pressure model, (b) symmetric two-pressure model, (c) thin-ship model for a standard Wigley hull, (d) Hogner model for a standard Wigley hull and (e) experimental signal.  $F =0.287$, $\sigma=0.1729$, $\ell=1.0167$, $\epsilon = 0.0106$, $\beta=0.1$, $\delta=0.0667$, $\nu$ is $1$\% of $\delta$ for the thin-ship model and $0$ for the Hogner model. The signal is taken at nondimensional distance $y_{s} = 2$. This Froude number corresponds to the standard Wigley hull travelling at $1.10\mathrm{ms}^{-1}$. The spectrogram in (f) is taken from a further experiment with a small hull of length $0.3\mathrm{m}$ travelling at $0.49\mathrm{ms}^{-1}$, leading to the slightly different Froude number $F =0.286$, where the signal is taken at $y_s=2.67$.  White and red arrows indicate effects of wave interference at the fold and sliding frequency mode, respectively.}
		\label{fig:experimental_low_froude_comparison}
	\end{figure}

\begin{figure}
		\centering
		\begin{tabular}{ll}
			(a) Single-pressure model & (b) Symmetric two-pressure model \\
			\includegraphics[scale=1]{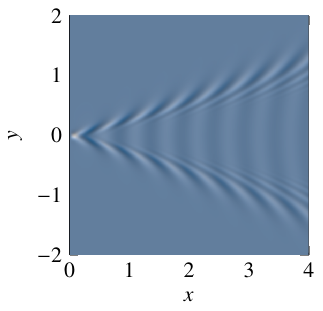} &
			\includegraphics[scale=1]{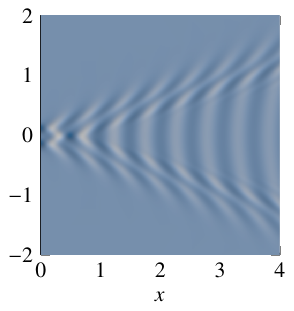} \\
			(a) Michell's thin-ship model & (b) Hogner model \\
			\includegraphics[scale=1]{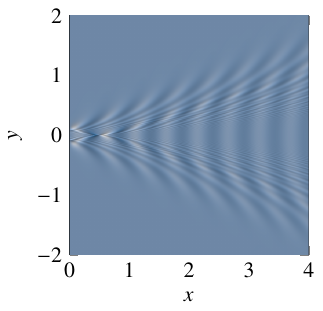} &
			\includegraphics[scale=1]{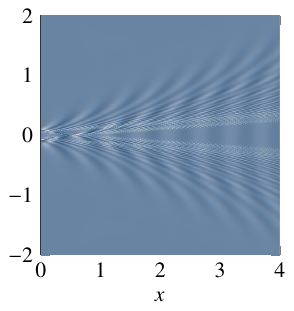}
		\end{tabular}
		\caption{Surface plots for the four models with $F = 0.287$, $\beta=0.1$, $\delta=0.0667$, $\sigma=0.1729$, $\ell=1.0167$, $\epsilon = 0.0106$, $\nu$ is $1$\% of $\delta$ for the thin-ship model and $0$ for the Hogner model. Note that we are only computing the far-field component of the solution. All of the surfaces are plotted on the same discrete grid.}
		\label{fig:just_surfaces}
	\end{figure}

	Figure~\ref{fig:experimental_low_froude_comparison}(a)-(d) shows wave elevation signals and spectrograms for our four models for the specific Froude number $F=0.287$.  The other parameter values are the same as those used in figure~\ref{fig:amplitudes}(a).  The spectrograms in panels (e)-(f) are from experimental data and will explained in \S~\ref{sec:experimental}.  For this relatively low Froude number, the transverse waves are visible in the wave train (see figure~\ref{fig:just_surfaces}) and in the spectrograms there is more energy (colour intensity) along the transverse branch than the divergent branch.  Further, the four wave amplitude functions in figure~\ref{fig:amplitudes}(a) are very close to each other for small values of $\psi$, which explains why the transverse branches in each of figure~\ref{fig:experimental_low_froude_comparison}(a)-(d) are all very similar in their colour intensity.  Turning to the folds of the spectrograms (where the constant frequency and sliding frequency branches meet), which correspond to values of $\psi$ near $\psi_c$, we see that the spectrogram for the symmetric two-pressure model appears to replicate that for Michell's thin ship theory very well, including the prominent area of low intensity (blue spot) situated roughly in the centre of the fold (indicated by the white arrow).  This blue spot is due to the zeros of the wave amplitude functions near $\psi=\psi_c$, caused by destructive wave interference in the divergent waves near the Kelvin wedge angle $\theta=\theta_{\mathrm{wedge}}$.  Note the spectrogram for the single pressure does not exhibit this wave interference, which is a clear deficiency in this basic model.
	
	The final observation we make about figure~\ref{fig:experimental_low_froude_comparison}(a)-(d) is that the sliding frequency mode in the spectrograms for the Michell (c) and Hogner (d) models are clearly present, even for high frequencies, including a number of low intensity regions (blue spots) due to subsequent zeros of the wave amplitude functions (indicated by red arrows; also noted by \cite{liang2024}).  These high frequency components would be damped out by viscosity, which is not included in our model, and, as such, can effectively be ignored.  The minimal models do not show the full sliding frequency modes as their wave amplitude functions decay quickly to zero for larger values of $\psi$.  
	
	At this stage it is worth plotting images of the wave patterns themselves for these four models.  These are provided in figure~\ref{fig:just_surfaces}(a)-(d) using the same parameters as in figure~\ref{fig:experimental_low_froude_comparison}(a)-(d).  One observation is that in each of (b)-(d) there is wavemaking at the bow and stern of the disturbance, whether that be for the symmetric two-pressure model in (b) or the more sophisticated models for the Wigley hull in (c) and (d).  On the other hand, the simple single-pressure model in (a) gives rise to a cleaner wave pattern without the complicated consequences of wave interference.  Another important observation is that it is very difficult to even qualitatively predict or describe the properties of these wave patterns in the time-frequency domain simply from viewing these images of wave patterns. Further, it is certainly not possible to judge how well the symmetric two-pressure model mimics the more sophisticated models in terms of wave interference from these images.  It is for this reason that we argue that the appropriate tools for our study are the wave amplitude function and the spectrogram.
	
	Another point to make about the images in figure~\ref{fig:just_surfaces} is that they take very different times to compute.  The single pressure (a) and symmetric two-pressure model (b) took roughly 20 seconds to generate while the thin-ship (c) and Hogner models (d) took roughly 20 minutes and 2 hours, respectively.  These significant differences provide one of the main motivations for exploring the basic minimal models in more detail with the goal of mimicking key features of the more complicated models without the computational expense.
	
	\begin{figure}
		\centering
		\begin{tabular}{ll}
			(a) Single-pressure model  & (b) Symmetric two-pressure model  \\
			\includegraphics[scale=1]{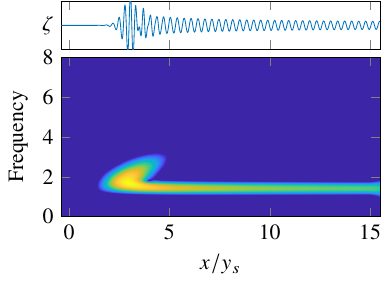} &
			\includegraphics[scale=1]{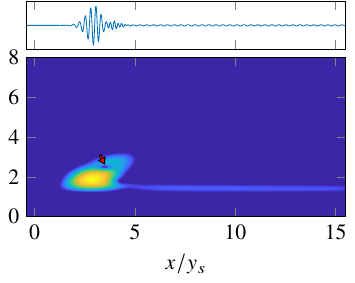} \\
			(c) Michell's thin-ship model & (d) Hogner model \\
			\includegraphics[scale=1]{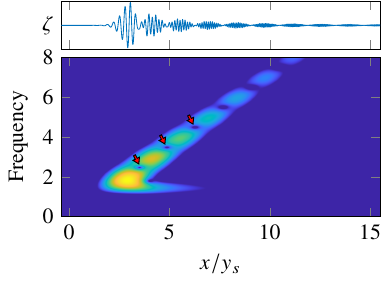} &
			\includegraphics[scale=1]{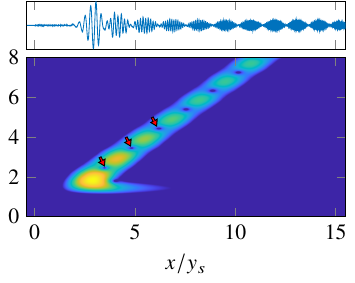} \\
			(e) Experimental standard hull size & (f) Experimental small hull \\
			\includegraphics[scale=1]{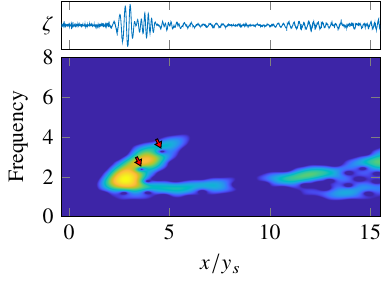} &
			\includegraphics[scale=1]{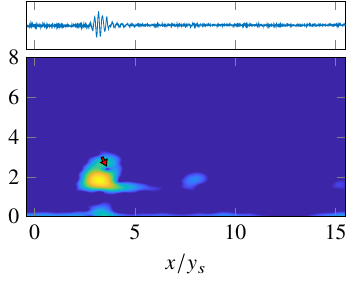}
		\end{tabular}
		\caption{Spectrograms for the (a) single-pressure model, (b) symmetric two-pressure model, (c) thin-ship model for a standard Wigley hull, (d) Hogner model for a standard Wigley hull and (e) experimental signal.  $F =0.334$, $\sigma=0.3086$, $\ell=1.0167$, $\epsilon = 0.0103$, $\beta=0.1$, $\delta=0.0667$, $\nu$ is $1$\% of $\delta$ for the thin-ship model and $0$ for the Hogner model. The signal is taken at $y_{s} = 2$. This Froude number corresponds to the standard Wigley hull of length $1.5\mathrm{m}$ travelling at $1.28\mathrm{ms}^{-1}$. This speed was chosen to minimise the transverse waves along the centreline according to Michell's thin-ship model. The spectrogram in (f) is taken from a further experiment with a small hull of length $0.3\mathrm{m}$ travelling at $0.57\mathrm{ms}^{-1}$, leading to the slightly different Froude number $F =0.332$, where the signal is taken at $y_s=2.67$.}
		\label{fig:experimental_interesting_froude_comparison}
	\end{figure}
	
\begin{figure}
		\centering
		\begin{tabular}{ll}
			(a) Single-pressure model & (b) Symmetric two-pressure model \\
			\includegraphics[scale=1]{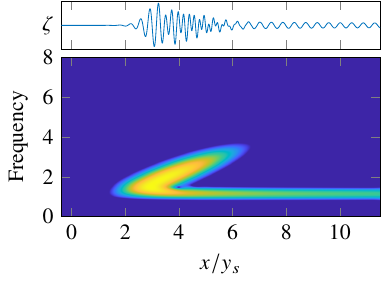} &
			\includegraphics[scale=1]{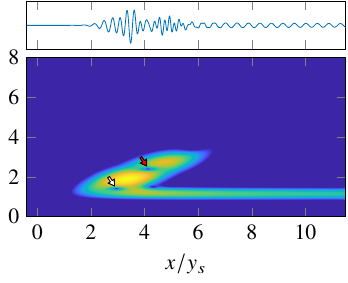} \\
			(c) Michell's thin-ship model & (d) Hogner model \\
			\includegraphics[scale=1]{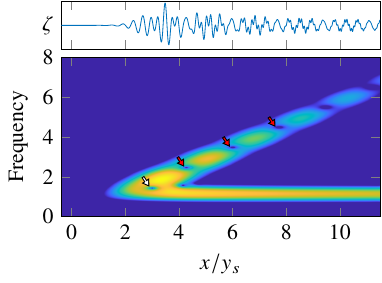} &
			\includegraphics[scale=1]{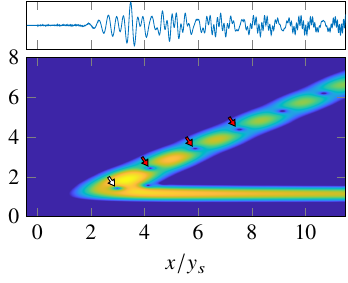} \\
			(e) Experimental standard hull size & (f) Experimental small hull  \\
			\includegraphics[scale=1]{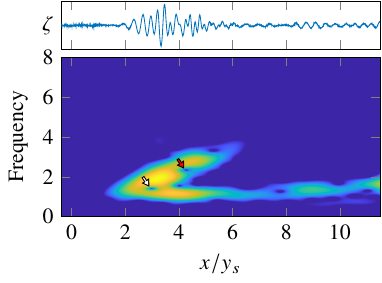} &
			\includegraphics[scale=1]{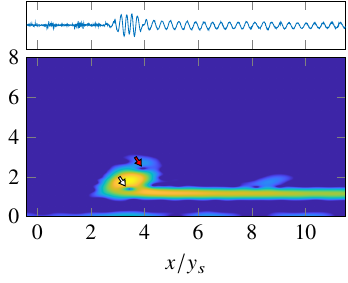} \\
		\end{tabular}
		\caption{Spectrograms for the (a) single-pressure model, (b) two-pressure model, (c) thin-ship model for a standard Wigley hull, (d) Hogner model for a standard Wigley hull and (e) experimental signal. The experimental signal was recorded in time and then converted to distance and nondimensionalised. $F=0.370$, $\sigma=0.1915$, $\ell=1.0487$, $\epsilon = 0.0188$, $\beta=0.1$, $\delta=0.0667$, $\nu$ is $1$\% of $\delta$ for the thin-ship model and $0$ for the Hogner model. The signal is taken at $y_{s} = 2$. This Froude number corresponds to the standard Wigley hull travelling at $1.42\mathrm{ms}^{-1}$. The spectrogram in (f) is taken from a further experiment with a small hull of length $0.3\mathrm{m}$ travelling at $0.63\mathrm{ms}^{-1}$, leading to the slightly different Froude number $F =0.367$, where the signal is taken at $y_s=2.67$.}
		\label{fig:experimental_high_froude_comparison}
	\end{figure}

	To illustrate one of the challenges with our approach, we show further spectrograms calculated from our minimal and Michell models in figures~\ref{fig:experimental_interesting_froude_comparison} and \ref{fig:experimental_high_froude_comparison}.  These are analogous to those presented in figure~\ref{fig:experimental_low_froude_comparison} (which was for $F=0.287$), except that now the Froude number is $F=0.334$ in figure~\ref{fig:experimental_interesting_froude_comparison} and $0.370$ in figure~\ref{fig:experimental_high_froude_comparison}.  In each case we have followed the algorithm outlined in \S\ref{sec:pressure_estimation} to choose the parameters in the symmetric two-pressure model.  The value $F=0.334$ is special because, for the Wigley hull we are concerned with here, this Froude number gives $A(0)=0$ (see the blue curve in figure~\ref{fig:amplitudes}(b), which clearly decays to zero as $\psi$ decreases to $\psi=0$).  That is, for this special speed, Michell's thin ship theory suggests that destructive interference of transverse waves is so pronounced that the stationary-phase approximation from \S\ref{sec:statphase} predicts the centreline is flat.  As a consequence, for this rather special case, the spectrograms in figure~\ref{fig:experimental_interesting_froude_comparison} (c)-(d) do not show the high intensity constant frequency mode like those in figure~\ref{fig:experimental_low_froude_comparison}.  On the other hand, the match between the symmetric two-pressure model and Michell's thin ship theory in figure~\ref{fig:amplitudes}(b) for larger values of $\psi$ is essentially as good as it is in figure~\ref{fig:amplitudes}(a).  Therefore the spectrogram in figure~\ref{fig:experimental_interesting_froude_comparison} (b) for $F=0.334$ shows a strong match with that in figure~\ref{fig:experimental_interesting_froude_comparison} (c) around the fold, corresponding to the important divergent waves near the boundary of the Kelvin wedge.
	
	Very briefly, we include in figure~\ref{fig:experimental_high_froude_comparison} spectrograms drawn for $F=0.370$.  Here the Froude number is slightly higher than in figure~\ref{fig:experimental_interesting_froude_comparison} (which was for $F=0.334$), but the qualitative behaviour of the constant frequency branch is closer to
	figure~\ref{fig:experimental_low_froude_comparison} (which was for $F=0.287$).  The reason is that this slightly higher hull speed is enough to avoid the exceptional destructive wave interference of transverse waves.  In fact, the value $F=0.370$ in figure~\ref{fig:experimental_high_froude_comparison} is special because now the wave amplitude function for the Michell model is zero at the critical angle $\psi_c$ (see the blue curve in figure~\ref{fig:amplitudes}(c)).  This destructive interference of divergent waves gives rise to an area of low intensity (blue spot) right in the middle of the fold of the spectrogram in figure~\ref{fig:experimental_high_froude_comparison}(c).  Encouragingly, our minimal model with two symmetric pressures is able to capture this rather complicated behaviour in the spectrogram very well (see figure~\ref{fig:experimental_high_froude_comparison}(b)).
	
	\section{Results for transom-stern hull}\label{sec:transom}
	
	
	In this section we turn to our second hull shape, the Wigley transom-stern hull (see figure~\ref{fig:bodyline_diagrams}(b)). The motivation for this hull is that, while it is thin for the most part, the stern is a flat (vertical) face that is perpendicular to the centreline. As such, a straightforward application of Michell's thin ship theory is not at all appropriate, as this model requires the hull to be represented by $y=Y(x,z)$ for all $x$ and that $Y_x$ is small, whereas the stern of the Wigley transom-stern hull is described by $x=$ constant (crudely speaking, $Y_x$ is infinite there).
While the Hogner model can handle the transom hull by relating the strength of the distribution of sources to the $x$-component of the normal to the hull surface, thereby allowing for the vertical transom stern (or any stern), this approach is still deficient in that it does not allow for flow separation that is caused by the transom stern.

For these reasons, when considering the transom-stern hull, we apply Michell's thin ship theory and the Hogner model to an adjusted hull shape which includes a virtual appendage that acts to enclose separation zone.  For the minimal model, we choose is the general two-pressure model \eqref{eq:two_pressure_transom} from \S\ref{sec:miminal}. We attempt to select parameters in this minimal model to mimic key features of the transom-stern hull in the time-frequency domain as predicted by the Hogner model. We note that the front half of the transom stern hull, $x<0$, is the same as the symmetric hull; therefore, we use the same parameter values $\epsilon_1$, $\sigma_1$ and $\ell_1$ in the {two-pressure virtual} model as we used in the {symmetric} two-pressure model.

	{To determine the remaining parameters \(\epsilon_{2}\), \(\sigma_{2}\) and \(\ell_{2}\), we manually adjusted them while comparing the two amplitude functions. We first increased \(\ell_{2}\) to move the centre of the pressure to be closer to the end of the virtual appendage with the aim of aligning the zero amplitude locations of the two-pressure model with the local minima of the Hogner model. The two-pressure model is still symmetric at the point, though no longer about the midship. We then adjust the strength of the rear pressure, which breaks the symmetry of the model, meaning there is no longer exact cancellation between the waves generated at the bow and stern leading to local minima in the amplitude function. We adjust the strength of the stern pressure until the amplitude of the first local minimum matches between the two-pressure and Hogner models.}
	
\begin{figure}
		\centering
		\includegraphics[scale=1]{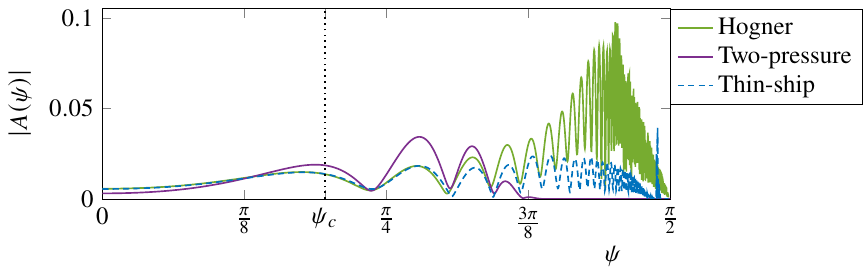}
		\caption{Magnitude of the wave amplitude functions for:
			the general two-pressure model \(\left. A_{2} \right|_{T}\) \eqref{eq:two_pressure_transom} with \(\epsilon_{1} = 0.010\), \(\sigma_{1} = 0.178\), \(\ell_{1} = 1.0167\), \(\epsilon_{2} = 0.014\), \(\sigma_{2} = 0.178\), \(\ell_{2} = 1.55\);
			Michell's thin ship model
			$\left.A_{M}\right|_{T}$
			({\color{colour1}blue})
			and the Hogner model
			$\left.A_{H}\right|_{T}$
			({\color{colour5}green}) for the Wigley transom-stern hull \eqref{eq:Wigley_transom_stern} with $\beta = 0.1$, $\delta=0.067$ and $\nu = 0.01\delta$, again with $F = 0.282$.}
		\label{fig:amplitudetransom}
	\end{figure}

	An example of such a fitting exercise is shown in figure~\ref{fig:amplitudetransom}. Here the Froude number is $F=0.282$. In this figure we see the wave amplitude function for the Hogner model in green does not have any roots but its magnitude does have a few obvious local maxima and minima for $\psi<3\pi/8$. Using our carefully chosen parameters for the general two-pressure model, the amplitude function in purple provides quite a good match with the Hogner model over this interval.  There is no match between the green and purple curves for larger values of $\psi$ but, as we have already argued, these high frequency waves will be damped out in practice and are therefore not important. We also see that the Michell model closely matches the Hogner model, although the locations of the local maxima and minima are slightly shifted. As mentioned above, the Michell model amplitude function is not able to be computed exactly, thus it has very little advantage over the Hogner model when modelling transom sterns using virtual appendages.
	
\begin{figure}
		\centering
		\begin{tabular}{ll}
			(a) General two-pressure model & (b) Michell's thin-ship model \\
			\includegraphics[scale=1]{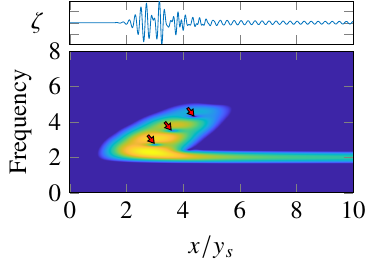} &
			\includegraphics[scale=1]{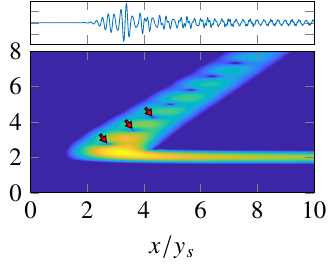} \\
			(c) Hogner model & (d) Experimental standard transom-stern hull \\
			\includegraphics[scale=1]{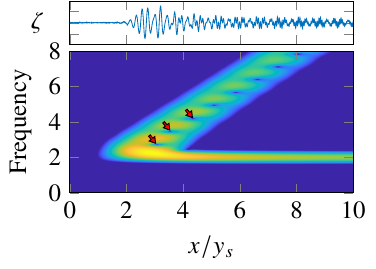} &
			\includegraphics[scale=1]{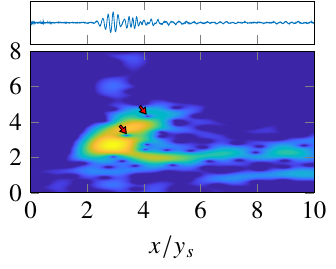}
		\end{tabular}
		\caption{Spectrograms for the (a) two-pressure model, with \(\epsilon_{1} = 0.010\), \(\sigma_{1} = 0.178\), \(\ell_{1} = 1.0167\), \(\epsilon_{2} = 0.014\), \(\sigma_{2} = 0.178\), \(\ell_{2} = 1.55\); (b) thin-ship model; (c) Hogner model and (d) experimental signal. For (b) and (c) the Wigley transom-stern hull is used. $F = 0.282$, $\beta=0.1$, $\delta=0.0667$, $\nu=0.01\delta$. The signals are taken at $y_{s} = 2$. The Froude number $F = 0.282$ is equivalent to the Wigley transom-stern hull, AMC 97-03, travelling at $1.08\mathrm{ms}^{-1}$.}
		\label{fig:transom_stern_comparison}
	\end{figure}

	Figure~\ref{fig:transom_stern_comparison}(a)-(c) shows the wave signals and spectrograms for the different transom-stern models for the Froude number $F=0.282$.
	The spectrogram for the {general two-pressure} model (a) agrees extremely well with the Hogner model (c) all the way up to the high frequency region that is not physically important, including low intensity blue spots (indicated by the red arrows) along the divergent branch caused by wave interference.
	It is clear that the experimental signal (d) also  exhibits wave cancellation resulting in the same blue spots. {Michell's thin-ship model (b) using the virtual appendage also exhibits the low intensity blue spots. Comparing to the experimental spectrogram, we can see the locations are on the wrong side of the divergent branch, indicating that the thin-ship model predicts the waves being in different phase. The Hogner model matches the location with the experimental results, including being on the correct side.}
	
	Finally, in terms of a visual representation of the
	{general two-pressure} model, we show in figure~\ref{fig:pressure_source_locations}(b) the location of the centre of the
	{two pressures} for the case $F=0.287$.
	
	\section{Experimental results}\label{sec:experimental}
	
	\subsection{Standard experimental hull size}
	
	Experiments were performed at the Australian Maritime College model test basin, which is a towing tank that is $35\mathrm{m}$ long and $12\mathrm{m}$ wide.  The left wall is $4\mathrm{m}$ from the sailing line. The two standard hull models used in these experiments were the AMC 93-02 Wigley hull and the AMC 97-03 hull, the latter of which is a Wigley hull from bow to the centre of the hull and rear half is a parallel-block with a transom stern.  Both hulls were $1.5\mathrm{m}$ long, which is at the smaller end of the scale for standard model hulls used at the Australian Maritime College.  In each case the beam is $0.15\mathrm{m}$ ($\beta=0.1$) and the draft is $0.1\mathrm{m}$ ($\delta=0.0667$).  A sensor recording the elevation of the surface was located $3\mathrm{m}$ from the sailing line and $10.5\mathrm{m}$ from the end of the tank.  This wave height gauge associates differences in capacitance with surface elevation.
	The location of the sensor was chosen to be far enough away from the hull to be considered just the far-field waves but not so far that the reflected waves from the side of the basin reach the sensor before the first group of waves have finished passing.  Further details of the experiments are explained in \cite{Buttle2020}.
	
	These experimental signals were recorded as functions of time and have been converted to functions of distance and then nondimensionalised for comparison with our numerical models. Three speeds for the experiments were chosen to exhibit interesting behaviours in the signals, as predicted by the wave amplitude function of Michell's thin ship theory: the speed $U=1.1\mathrm{ms}^{-1}$ is associated with a representative low Froude number ($F=0.287$);  the speed $1.28\mathrm{ms}^{-1}$ ($F=0.334$) corresponds to a wave amplitude function with a root at $\psi=0$; and the speed $1.42\mathrm{ms}^{-1}$ ($F=0.370$) was chosen to have zero amplitude at the angle $\psi=\psi_{c}$.
	
	Spectrograms computed from experimental data for the AMC 93-02 Wigley hull at $F=0.287$, $0.334$ and $0.370$ are shown in figures~\ref{fig:experimental_low_froude_comparison}(e), \ref{fig:experimental_interesting_froude_comparison}(e) and \ref{fig:experimental_high_froude_comparison}(e), respectively.  While these images are not as clean as the theoretical counterparts, there is remarkably strong agreement between the experimental spectrograms and the ones for both Michell's thin ship theory and the two-pressure model around the fold where the constant and sliding frequency modes meet, including the crucial regions of low intensity (blue spots) that are due to wave interference.  On the other hand, wave damping in the experiments lead to the spectrograms not showing high intensity regions away from fold, and so in this region the similarities are not so obvious.  There are also hints of wave reflection off the walls for large $x/y_s$ and lower frequency in figures~\ref{fig:experimental_low_froude_comparison}(e), \ref{fig:experimental_interesting_froude_comparison}(e) and \ref{fig:experimental_high_froude_comparison}(e).  Overall, these comparisons provide encouraging evidence that our theoretical study has genuine consequences for real physical models, and supports our argument that the two-pressure model can be viewed as a useful simple proxy for more complicated models, provided the parameters are chosen carefully.
	
	Similar comments can be made regarding the experimental spectrogram in figure~\ref{fig:transom_stern_comparison}(d) for the representative Froude number $F=0.282$.  There is some correspondence between the experimental spectrogram and the corresponding spectrogram for the Hogner model in (c), especially near the fold, while again there are differences away from the fold.  In (d) the experimental spectrogram includes a clear region of high intensity for dimensionless frequency roughly one and $6\lessapprox x/y_s < 10$, which is mostly a consequence of waves being reflected
	off the tank wall.  Obviously this feature is not observed in the theoretical models which assume an infinite domain.
	
\begin{figure}
		\centering
		\includegraphics[scale=1]{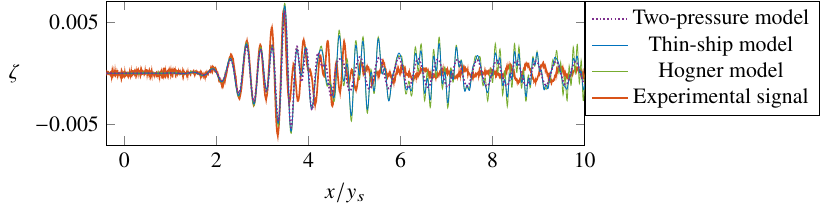}
		\caption{Comparison of the nondimensional experimental signal, orange, with the signals generated by the two-pressure, purple dashed, and thin-ship, blue, and Hogner, green, models. All signals are taken at nondimensional distance $y_{s}=2$. These data are for $F = 0.3704$, which corresponds to a standard Wigley hull of length $1.5\mathrm{m}$ travelling at $1.42\mathrm{ms}^{-1}$.}
		\label{fig:signal_comparison}
	\end{figure}

	In addition to comparing spectrograms, figure~\ref{fig:signal_comparison} shows the surface height signal from the AMC 93-02 Wigley experiment compared to the numerical models. The first few wavelengths are almost identical in phase and magnitude between the experiments and the numerical models. The poorer phase match later in the signal can be attributed to the waves being damped in the water in the experiment, while there are no damping effects in the numerical models.  Another reason that the experimental signal decays around $x/y_{s} = 6$ and the numerical models do not is that the experiments required the hull to start its motion from a stationary position in the test basin, while the numerical models assume the hull has been moving at constant speed for infinite time and distance before passing the sensor.  This is one of the reasons that the transverse branch in the experimental spectrogram appears much shorter in figures~\ref{fig:experimental_low_froude_comparison}(e), \ref{fig:experimental_interesting_froude_comparison}(e) and \ref{fig:experimental_high_froude_comparison}(e)
	figure~\ref{fig:transom_stern_comparison}(d)
	than the transverse branch in the numerical spectrograms.
	Indeed, the accelerating hull in the experiments also appears to affect the transverse branch in the spectrograms by making them slightly increasing instead of remaining constant in frequency.  Using time-dependent numerical models that take into account the experimental hull's acceleration, rather than models that assume steadily moving hulls, could account for this shorter and slightly increasing transverse branch \citep{Pethiyagoda2021}.
	
	\subsection{Smaller-scale experimental hulls}
	
	To provide supporting experimental results, we repeated the experiments with the same Wigley hull as in figure~\ref{fig:bodyline_diagrams}(a) and equation (\ref{eq:Wigley}), with $\beta=0.1$ and $\delta=0.067$, except the actual dimensions were $1/5$ of the standard-size hull described in the previous subsection.  That is, the dimensional length for this smaller-scale hull was $0.3\mathrm{m}$, while the beam and draft were
	$0.03\mathrm{m}$ and $0.02\mathrm{m}$, respectively.
	
	The motivation for using this very small hull was twofold,  First, we wanted to minimise any wave reflection off the walls, given the fixed size of the model test basin.  We can see the consequences of wave reflection in the spectrograms for the standard sized hull
	in figures~\ref{fig:experimental_low_froude_comparison}(e), \ref{fig:experimental_interesting_froude_comparison}(e) and \ref{fig:experimental_high_froude_comparison}(e).  With a smaller scale hull, we were able to delay this reflection until a much longer dimensionless time.  Second, another consequence of using the standard size hull is that we needed to accelerate the hull up to the cruising speed rather quickly.  The use of a smaller-scale hull was designed to overcome this deficiency.
	
	The smaller hull itself was manufactured using 3D printing with standard polylactic acid.  Another change to the experimental set up was that the wave height gauge (the sensor), which was located at $0.8\mathrm{m}$ from the sailing line for the smaller hull, was specially designed to detect the smaller amplitude signals.  To match the three Froude numbers used with the standard hull, we need to reduce the cruising speeds to be $0.49\mathrm{ms}^{-1}$, $0.57\mathrm{ms}^{-1}$ and $0.63\mathrm{ms}^{-1}$, respectively.  If we keep track of the third decimal place, then actually the resulting Froude numbers of $F=0.286$, $0.332$ and $0.367$ are very slightly different to the ones used for the standard sized hull, although we expect these differences to be negligible.
	
	Spectrograms for signals measured with this smaller-scale Wigley hull are shown in figures~\ref{fig:experimental_low_froude_comparison}(f), \ref{fig:experimental_interesting_froude_comparison}(f) and \ref{fig:experimental_high_froude_comparison}(f).  Broadly speaking, they compare well with the spectrograms for the standard sized hull, especially near the fold (where the constant frequency and sliding frequency modes meet).  As expected, the transverse branches of the spectrograms in figures ~\ref{fig:experimental_low_froude_comparison}(f) and \ref{fig:experimental_high_froude_comparison}(f) are much cleaner than with the standard sized hulls, as the unwanted effects of reflection and acceleration were no longer present.  Remarkably, the spectrogram in \ref{fig:experimental_interesting_froude_comparison}(f) does not include any significant high intensity along the transverse branch, which is consistent with the specially chosen Froude number that maximises destructive transverse wave interference according to Michell's thin ship model.  Another observation is that the higher frequency contributions along the divergent branch are not present with this small-scale experimental model, which is as expected given the extra challenges associated with detecting these very small wavelengths.
	
	\section{Discussion}\label{sec:discussion}
	
	\begin{table}
		\centering
		\begin{tabular}{cccccc}
			\toprule
			Model &
			\begin{tabular}[c]{@{}c@{}}Physically \\ realistic\\ parameters\end{tabular}
			&
			\begin{tabular}[c]{@{}c@{}}Mathematically \\ convenient\\ parameters\end{tabular}
			&
			\begin{tabular}[c]{@{}c@{}}Analytic\\ formula\end{tabular}
			&
			\begin{tabular}[c]{@{}c@{}}Describes\\ wave \\ cancellation\end{tabular}
			&
			Runtime \\ \midrule
			Hogner
			&
			\begin{tabular}[c]{@{}c@{}}hull geometry, \\ speed\end{tabular}
			&
			--
			&
			no
			&
			yes
			&
			\begin{tabular}[c]{@{}c@{}}hundreds of \\ minutes\end{tabular}
			\\
			Thin-ship
			&
			\begin{tabular}[c]{@{}c@{}}hull geometry, \\ speed\end{tabular}
			&
			$\nu$ & yes & yes
			&
			\begin{tabular}[c]{@{}c@{}}tens of \\ minutes\end{tabular}
			\\
			Two pressure
			&
			\begin{tabular}[c]{@{}c@{}}length scale, \\ speed\end{tabular}
			&
			\begin{tabular}[c]{@{}c@{}}pressure strength,\\ half-width, distance\end{tabular}
			&
			yes & yes &
			\begin{tabular}[c]{@{}c@{}}tens of \\ seconds\end{tabular}
			\\
			Single pressure & speed
			&
			\begin{tabular}[c]{@{}c@{}}pressure strength,\\ half-width\end{tabular}
			&
			yes & no & seconds \\
			\bottomrule
		\end{tabular}
		\caption{Various properties of the different mathematical models considered for the bow-stern symmetric hull in this study.
		}
		\label{table1}
	\end{table}
	
	This study is concerned with exploring to what extent minimal models with applied pressure distributions can be used to reproduce important features of ship wakes made by thin hulls moving steadily through still water, especially in terms of mimicking the time-frequency maps generated by surface wave elevation signals measured at a single point in space as the hull travels by. Part of the motivation for this work arises from the extensive use of various types of such simplified models in the physics and applied mathematics literature in the absence of a strong connection with more sophisticated models that explicitly take into account features of ship hulls \citep{colen21,Darmon2014,Ellingsen2014,GabrielDavid2017,Li2002,Li2016,lo21,Miao2015,Parau2002,Pethiyagoda2015,Pethiyagoda2021a,Rabaud2014,Scullen2011,Smeltzer2019}.  Further motivation comes from the recent use of spectrograms \citep{Didenkulova2013,Liang2022,Luo2022,Pethiyagoda2017,Pethiyagoda2018,Pethiyagoda2021,Ratsep2020,Ratsep2020b,Ratsep2021,Torsvik2015} to study how time-frequency information encodes properties of ship hulls that cause waves in the first place.
	
	For a standard Wigley hull, we have applied Michell's thin ship theory and used the corresponding wave amplitude function in \eqref{eq:thin_ship_amplitude} to match with a minimal model that involves two identical Gaussian pressure distributions applied to the surface, one located near the bow of the hull, and the other located near the stern.  We also consider a more simplistic model with a single pressure distribution and a further model, the Hogner model, which is more complicated than Michell's theory.  To illustrate our results, we chose three ship speeds, including one that gives rise to maximum destructive interference in the transverse waves along the centreline (at least according to Michell's thin ship theory) as well as another example where maximum destructive interference of waves at the caustic.
	We generated spectrograms in all cases to demonstrate shared features and crucial differences between the models.  In addition, for a transom-stern hull, we explore the challenges that arise in attempting to follow the same matching process, in part because both Michell's thin ship theory and the Hogner model have clear deficiencies for a transom-stern hull.  In order to mimic the flow separation that occurs due to the steep geometry of a transom stern, we extend the definition of the hull to include a virtual appendage, and apply the matching process to the extended hull.  Our theoretical study is supported by experimental data taken from steadily moving Wigley and transom-stern hull in a model test basin.
	
	We now summarise the main conclusions of our study.  For a standard Wigley hull, both Michell's thin ship theory and the Hogner model work well in terms of explicitly incorporating details of the hull surface without the challenge of solving for the details of the waterline (for detailed discussions about including waterline calculations in a boundary integral framework, see \cite{He2018,He2021}, for example), provided the hull itself is sufficiently thin \citep{Zhu2017a}.  Further, these two models give rise to similar wave amplitude functions, except in the part of the domain near $\psi=\pi/2$ which is not physically important.
	In the context of studying wave responses in the time-frequency domain, Michell's thin ship theory has the advantage over the Hogner model in that the wave amplitude function $A(\psi)$ in (\ref{eq:stationary_phase_version}) can be written down exactly (see equations (\ref{eq:amplitudeM}) and (\ref{eq:thin_ship_amplitude})), greatly reducing the computational time and allowing for detailed analytical scrutiny.  These and other properties are summarised in table~\ref{table1}.
	
	Using the formula (\ref{eq:thin_ship_amplitude}), we are able to devise a methodology for choosing parameters in the symmetric two-pressure model which involves selecting the Gaussian half-width and separation of the pressure distributions to match a root and local maximum of $|A(\psi)|$ with Michell's thin ship theory.   The novelty in our approach is based on our motivation to study minimal models that mimic the key features of more complicated models in the time-frequency domain, in particular via spectrograms.  We are not motivated to mimic wave resistance, which is especially of interest for ship design and naval architecture, or to mimic transverse waves along the centreline, as these are not associated with the most distinct features of spectrograms.
	
	
	It is known from previous work that the most fundamental features of a spectrogram of a ship wake (namely the constant frequency branch, the sliding frequency branch and the fold connecting them) are reproduced by a very simple mathematical model with a single Gaussian pressure distribution applied to the surface \citep{Pethiyagoda2017,Pethiyagoda2018}.  On the other hand, we demonstrate here how at least two pressure distributions are required to produce the important interference patterns that appear along the high intensity regions of the spectrogram from either Michell's thin ship model or experimental data.  This minimum requirement was first hinted at the end of our previous study \citep{Pethiyagoda2018} and is broadly in the same spirit of other research involving two pressure distributions \citep{colen21,He2015,Li2018,Noblesse2014,Zhu2015}, although, importantly, those other studies did not involve detailed time-frequency analysis.
	
	For a Wigley hull that is not particularly thin (that is, if the dimensionless beam $\beta$ is not small), Michell's thin ship theory does not work as well as the Hogner model \citep{Zhu2017a}.  Thus for such hulls it would no longer be appropriate to attempt to match minimal models with the Michell model.  We do not pursue this issue further as we wish to concentrate on thin ship hulls in this study.

For a transom-stern hull, we apply the Michell and Hogner theories to an extended version of the hull that includes a virtual appendage designed to mimic flow separation at behind the vessel.  In this case we find that a reasonable minimal model involves two different pressure distribution.  The resulting spectrogram for this general two-pressure appears to replicate the key features of that for both the Michell and Hogner models very well, including the important interference effects.  This type of progress in studying minimal models would not be possible without the kind of time-frequency analysis outlined here.
	
	Experimental results with the standard-sized hulls serve to support our theoretical findings in the time-frequency domain with high fidelity data.  In particular, we have shown that our experimental spectrograms for the Wigley hull include low intensity regions (blue spots) at the fold where the two main branches meet and at the lower part of the sliding frequency mode, in the same manner as that predicted by both Michell's thin ship theory and the Hogner model.
	
	Regarding the complementary experiments with a smaller Wigley hull, these are useful as they provide an additional check and test of both the standard hull and the theoretical results.  In particular, they provide more reasonable results for the transverse branch of the spectrograms than the standard hull does, as the smaller scale of the hull in the same sized towing tank means that reflection off the walls and acceleration is greatly reduced.
	
	Finally, we acknowledge the two hulls (standard Wigley and Wigley transom-stern) used in this experimental study are not seaworthy and, for example, were not allowed to pitch or heave in the usual way.  Indeed, the hulls were fixed in place with zero degrees of freedom.  Instead, these hulls were chosen especially for their simple parameterisation and their wide use in the literature.  Further, the speeds we concentrated on in the towing tank were carefully chosen by first consulting with Michell's thin ship theory.  In particular, two of the speeds were determined by setting the cosine term in (\ref{eq:thin_ship_amplitude}) to be zero at $\psi=0$ and $\psi_c=\arctan(1/\sqrt{2})$ and solving for the corresponding Froude numbers, in order to test extreme cases of wave interference.  The decision to run further experiments with smaller-scaled hulls was made to minimise the unwanted effects from reflection and acceleration.  In this way, the design of the experiments (including the hull shapes and scale, cruising speeds, operating conditions, etc.) was driven to explicitly support and, in some cases, validate the theoretical and numerical study, not the other way around.  The question of how well our time-frequency analysis and the key conclusions extend to more realistic hulls and operating conditions is left for future consideration.

	\bibliographystyle{jfm}
	\bibliography{references}

	
\end{document}